\documentclass[sigconf]{acmart}

\usepackage{xcolor}
\usepackage{graphicx}
\usepackage{booktabs} 

\copyrightyear{2026}
\acmYear{2026}
\setcopyright{cc}
\setcctype{by-nc-nd}
\acmConference[CHI '26]{Proceedings of the 2026 CHI Conference on Human Factors in Computing Systems}{April 13--17, 2026}{Barcelona, Spain}
\acmBooktitle{Proceedings of the 2026 CHI Conference on Human Factors in Computing Systems (CHI '26), April 13--17, 2026, Barcelona, Spain}
\acmDOI{10.1145/3772318.3790606}
\acmISBN{979-8-4007-2278-3/2026/04}

\definecolor{changecolor}{HTML}{000000}
\newcommand{\change}[1]{\textcolor{changecolor}{#1}}

\begin{document}

\pdfstringdefDisableCommands{\def\\{ }}

\title[Grand Challenges around Designing Computers’ Control Over Our Bodies]{Grand Challenges around Designing Computers’ Control Over Our Bodies}


\author{Florian `Floyd' Mueller}
\email{floyd@exertiongameslab.org}
\orcid{0000-0001-6472-3476}
\affiliation{%
  \institution{Monash University}
  \city{Melbourne}
  \country{Australia}
}

\author{Nadia Bianchi-Berthouze}
\email{n.berthouze@ucl.ac.uk}
\orcid{0000-0001-8921-0044}
\affiliation{%
  \institution{University College London}
  \city{London}
  \country{United Kingdom}
}

\author{Misha Sra}
\email{sra@ucsb.edu}
\orcid{0000-0001-8154-8518}
\affiliation{%
  \institution{University of California}
  \city{Santa Barbara, CA}
  \country{United States}
}

\author{Mar Gonzalez-Franco}
\email{margon@google.com}
\orcid{0000-0001-6165-4495}
\affiliation{%
  \institution{Google}
  \city{Seattle, WA}
  \country{United States}
}

\author{Henning Pohl}
\email{henning@cs.aau.dk}
\orcid{0000-0002-1420-4309}
\affiliation{%
  \institution{Aalborg University}
  \city{Aalborg}
  \country{Denmark}
}

\author{Susanne Boll}
\email{susanne.boll@uol.de}
\orcid{0000-0003-4293-1623}
\affiliation{%
  \institution{University of Oldenburg}
  \city{Oldenburg}
  \country{Germany}
}

\author{Richard Byrne}
\email{rich@exertiongameslab.org}
\orcid{0000-0002-6391-0496}
\affiliation{%
  \institution{Monash University}
  \city{Melbourne}
  \country{Australia}
}

\author{Arthur Caetano}
\email{caetano@ucsb.edu}
\orcid{0000-0003-0207-5471}
\affiliation{%
  \institution{University of California}
  \city{Santa Barbara, CA}
  \country{United States}
}

\author{Masahiko Inami}
\email{drinami@star.rcast.u-tokyo.ac.jp}
\orcid{0000-0002-8652-0730}
\affiliation{%
  \institution{University of Tokyo}
  \city{Tokyo}
  \country{Japan}
}

\author{Jarrod Knibbe}
\email{j.knibbe@uq.edu.au}
\orcid{0000-0002-8844-8576}
\affiliation{%
  \institution{The University of Queensland}
  \city{St Lucia, Queensland}
  \country{Australia}
}

\author{Per Ola Kristensson}
\email{pok21@cam.ac.uk}
\orcid{0000-0002-7139-871X}
\affiliation{%
  \institution{University of Cambridge}
  \city{Cambridge}
  \country{United Kingdom}
}

\author{Xiang Li}
\email{xl529@cam.ac.uk}
\orcid{0000-0001-5529-071X}
\affiliation{%
  \institution{University of Cambridge}
  \city{Cambridge}
  \country{United Kingdom}
}

\author{Zhuying Li}
\email{zhuying9405@gmail.com}
\orcid{0000-0001-5474-6949}
\affiliation{%
  \institution{Southeast University}
  \city{Nanjing}
  \country{China}
}

\author{Joe Marshall}
\email{joe.marshall@nottingham.ac.uk}
\orcid{0000-0001-9666-786X}
\affiliation{%
  \institution{University of Nottingham}
  \city{Nottingham}
  \country{United Kingdom}
}

\author{Louise Petersen Matjeka}
\email{me@louisepmatjeka.com}
\orcid{0000-0002-9716-7207}
\affiliation{%
  \institution{NTNU}
  \city{Trondheim}
  \country{Norway}
}

\author{Minna Nygren}
\email{minna.nygren@ucl.ac.uk}
\orcid{0000-0002-4086-2479}
\affiliation{%
  \institution{University College London}
  \city{London}
  \country{United Kingdom}
}

\author{Rakesh Patibanda}
\email{rakesh@exertiongameslab.org}
\orcid{0000-0002-2501-9969}
\affiliation{%
  \institution{Monash University}
  \city{Melbourne}
  \country{Australia}
}

\author{Sara Price}
\email{sara.price@ucl.ac.uk}
\orcid{0000-0002-5092-1663}
\affiliation{%
  \institution{University College London}
  \city{London}
  \country{United Kingdom}
}

\author{Harald Reiterer}
\email{harald.reiterer@uni-konstanz.de}
\orcid{0000-0001-8528-8928}
\affiliation{%
  \institution{University of Konstanz}
  \city{Konstanz}
  \country{Germany}
}

\author{Aryan Saini}
\email{aryan@exertiongameslab.org}
\orcid{0000-0002-2844-3343}
\affiliation{%
  \institution{Monash University}
  \city{Melbourne}
  \country{Australia}
}

\author{Oliver Schneider}
\email{oliver.schneider@uwaterloo.ca}
\orcid{0000-0002-6240-8445}
\affiliation{%
  \institution{University of Waterloo}
  \city{Waterloo, ON}
  \country{Canada}
}

\author{Ambika Shahu}
\email{ambika.shahu@it-u.at}
\orcid{0000-0001-8753-4942}
\affiliation{%
  \institution{IT:U}
  \city{Linz}
  \country{Austria}
}


\author{Phoebe O. Toups Dugas}
\email{phoebe.toupsdugas@monash.edu}
\orcid{0000-0002-6192-2004}
\affiliation{%
  \institution{Monash University}
  \city{Melbourne}
  \country{Australia}
}

\author{Don Samitha Elvitigala}
\email{don.elvitigala@monash.edu}
\orcid{0000-0002-8013-5989}
\affiliation{%
  \institution{Monash University}
  \city{Melbourne}
  \country{Australia}
}

\renewcommand{\shortauthors}{Mueller, et al.}

\begin{abstract}
Advances in emerging technologies, such as on-body mechanical actuators and electrical muscle stimulation, have allowed computers to take control over our bodies. This presents opportunities as well as challenges, raising fundamental questions about agency and the role of our bodies when interacting with technology. To advance this research field as a whole, we brought together expert perspectives in a week-long seminar to articulate the grand challenges that should be tackled when it comes to the design of computers’ control over our bodies. These grand challenges span technical, design, user, and ethical aspects. By articulating these grand challenges, we aim to begin initiating a research agenda that positions bodily control not only as a technical feature but as a central, experiential, and ethical concern for future human–computer interaction endeavors.


\end{abstract}



\begin{teaserfigure}
  \includegraphics[width=\textwidth]{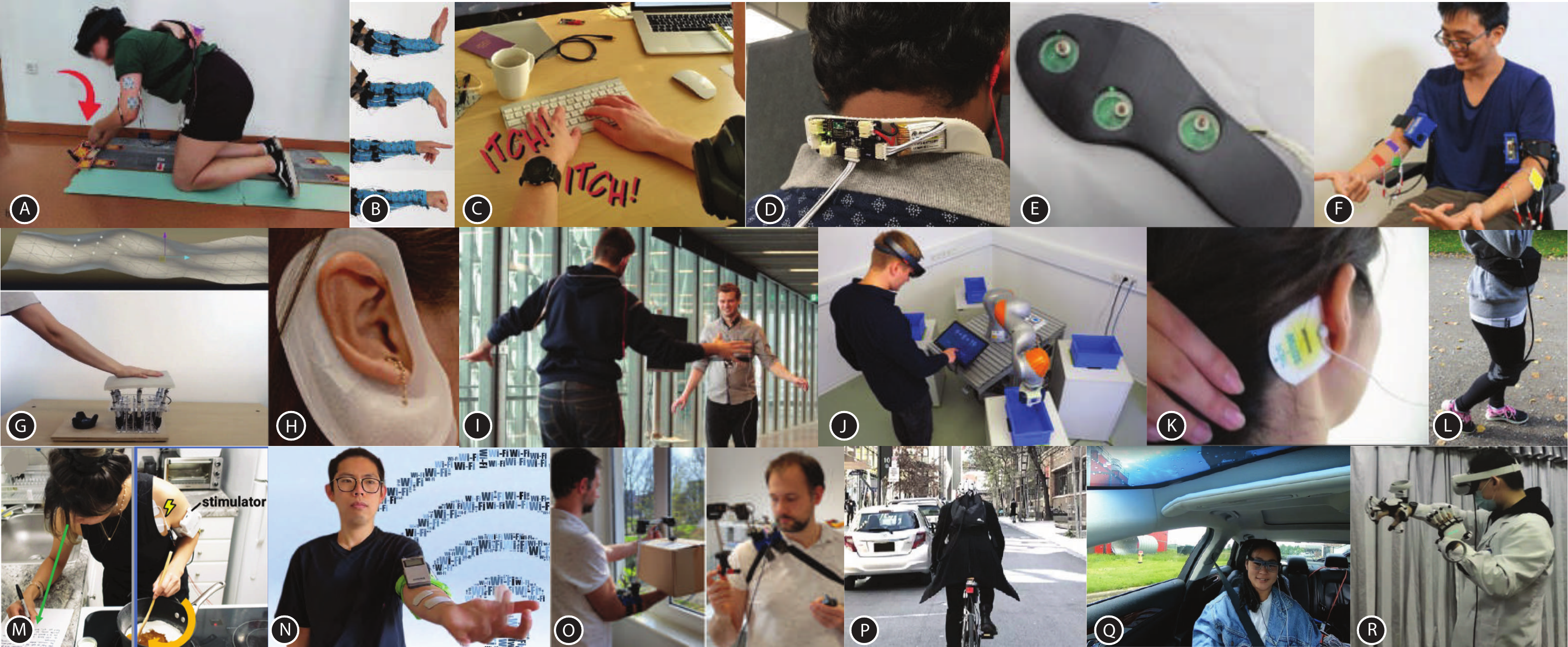}
  \caption{Example interfaces that take control over our bodies via (A) guiding motor actions \cite{Shahu2023Skillab}, (B) automatic calibration of high-density electrical muscle stimulation \cite{knibbe_automaticEMS_2017}, (C) itching sensations~\cite{Pohl2018Itching}, (D) proprioceptive feedback~\cite{Sra2019GVS}, (E) tickling sensations~\cite{Elvitigala2022}, (F) electrical muscle stimulation for gameplay~\cite{Patibanda2022Fused}, (G) shape displays providing encounter haptics \cite{steed2021mechatronic}, (H) pneumatic bodily extensions \cite{Saini_Pneuma_2024}, (I) surrendering a sense of balance \cite{byrne2016BalanceNinja}, (J) visualization of robot motion intent~\cite{Gruenefeld:2020}, (K) inhibited movement~\cite{Pohl2017InhibitMovement}, galvanic vestibular stimulation \cite{Maeda2005}, (L) compression feedback \cite{Pohl2017InhibitMovement}, (M) multitasking through computer-controlled muscle stimulation~\cite{nith2024Splitbody}, (N) allergy-like electrical muscle stimulations based on WiFi signals~\cite{Liu:Superpower:2024}, (O) wearable robotic limbs \cite{saberpour2023computational}, (P) eBikes ~\cite{andres2020introducing}, (Q) utilizing physiology to identify anxiety when control is given away, (R) force feedback gloves.}
  \Description{A collage of 12 images (labeled A-L) showcasing various interactive systems that involve shifting bodily control through different technologies: (A) shows a system guiding motor actions, with a person kneeling and reaching forward to represent guided movement; (B) demonstrates the automatic calibration of high-density electric muscle stimulation (EMS) through a sequence of arm movements; (C) features a setup inducing itching sensations on the hand using a tactile feedback device; (D) illustrates a system inducing proprioceptive feedback with a device placed on the back of a person’s neck; (E) shows a device inducing tickling sensations, presented as a close-up of an electronic sensor; (F) depicts a person with sensors on their arms watching computer-controlled body parts playing games; (G) shape displays are actuated and can provide encounter haptics when in contact with our bodies; (H) highlights pneumatic bodily extensions supporting embodied cognition, demonstrated with an ear-attached sensor patch; (I) shows an activity involving surrendering and regaining a sense of balance, with two people interacting indoors; (J) presents a system inhibiting freedom of movement, featuring a person using a touchscreen while interacting with a robotic arm; (K) depicts a close-up of a person's head with a sensor patch behind their ear, potentially monitoring or modulating sensory input;(L) shows a wearable device altering balance during outdoor walking, with sensors attached to the body; (M) shows a person drawing, while a computer controls their other hand to stir a curry using EMS; (N) features a person wearing EMS powered actuator to stimulate their hand based on the WiFi strength of the environment; (O) depicts a person wearing a customized robotic limb to lift extra weight and hold tools in peripersonal space; (P) shows a person cycling while wearing a EEG headset connected to the breaking system, giving away the control of a cycle to a computer based on EEG signals; (Q) shows a person wearing several physiology sensors while sitting inside an automated car; (R) demonstrates a person playing a game wearing a VR head set and force haptic gloves.}
  \label{fig:Teaser}
\end{teaserfigure}

\maketitle

\section{Introduction}

Traditional human-computer interaction (HCI) has largely assumed a \textit{controller-responder model} in which users command computers to execute tasks. In other words, interactions in which the user is in control. However, with shifts towards human-computer integration~\cite{Farooq2016} and human augmentation~\cite{Chignell2023}, this is often no longer the case.
Instead, systems are increasingly emerging where the computer takes control over users, acting on their behalf~\cite{oulasvirta2018}, influencing their behaviors~\cite{Adams2015}, or even actuating their bodies~\cite{Tamaki2011}.

Computers taking control over the user's body occurs across a broad range of application domains, with control taking on many forms, across different body locations, and ranging from subtle to forceful, enabled by a variety of technologies (see Figure \ref{fig:Teaser}). 
For example, in healthcare settings, computers control neuromuscular electrical stimulation (NMES) for physical therapy~\cite{Chattanooga-rehab}, while lower-limb exoskeletons help patients walk after surgery by dynamically adjusting control levels~\cite{CYBERDYNE}. 
In fitness settings, electrical muscle stimulation systems enhance athletic performance \cite{Compex}, while in industrial applications devices assist warehouse workers with lifting~\cite{SUITX} and reduce physical strain from repetitive tasks~\cite{PaexoSoftBack}, similar to military systems aiding soldiers with heavy loads~\cite{Dougherty_2024}. 
In entertainment settings, computers use subtle electric shocks to prompt players to release buttons \cite{ElectrocutionGame2021}, while the creative industry employs motor-controlled exoskeletons to support camera operators~\cite{SteadicamExovest}, and teleoperation systems control the operator's body to convey obstacles in remote locations~\cite{Shahu2023EmsObstacleAvoidance}.


While the overall challenges around human-computer integration have been explored previously (e.g., ~\cite{Mueller2020IntegrationNextSteps}), the specifics and consequences of computers taking control over our bodies appear to be under-articulated. This is surprising, as such a shift in control, while having significant potential to deepen our relationship with technology, raises significant  concerns and questions.
For example, how would a malfunction affect the user's safety? 
What measures ensure that the computer's actions match the user’s intentions and expectations? 
These examples highlight just a few of the issues that underline the importance of careful design and implementation of computers that take control over our bodies. 
Yet, our knowledge of how to design such experiences and the potential consequences of this kind of shift in control is limited.


In this article, we articulate a set of grand challenges around computers taking control over our bodies by investigating key issues and questions arising from this shift in control. 
This articulation was achieved by conducting a week-long seminar, where we brought expert perspectives together to comprehensively discuss the field. 
This approach takes inspiration from previous work that articulated grand challenges in other areas of HCI (e.g., WaterHCI, SportsHCI, human-food interaction and Human-Centered AI~\cite{Mueller2024WaterHCI, Elvitigala2024, Mueller2024FoodGC, Ozlem2023GrandChallengesHumanAI}).
With our articulation of grand challenges---with grand challenges defined as difficult but important problems that often require a time-frame of around 10~years to be solved \cite{Elvitigala2024}---we hope that we can begin to develop a structured research agenda that aids established researchers in their work and inspires PhD students to identify interesting research topics in order to move the field forward as a whole. 
With more and more systems emerging as a result of new actuation, sensing and AI technologies~\cite{Sun2022}, there is a timely need for this work.
By articulating grand challenges, we aim to facilitate an impactful and ethically responsible future for computers that take control over our bodies.

In the next section, we describe what we learned from prior work. 
Then, we articulate how we arrived at the grand challenges through our workshop, including how articulating different perspectives on control and motivations for yielding control helped with that.
We then articulate the grand challenges based on four categories (technology, design, user, and ethics) before presenting suggestions for future work.

\section{Related Work}

Research at the intersection of bodies and computers spans multiple traditions in HCI. We organize this work into three areas that directly informed the grand challenges articulated in this paper: (1) body-centric perspectives in HCI, (2) technologies that act on, or co-act with the human body, and (3) research on shared control, agency, and their experiential consequences. Together, these areas reveal both the opportunities and conceptual gaps that motivate the grand challenges outlined later.

\subsection{Body-Centric Perspectives in HCI}

Research on embodiment highlights how technologies shape bodily experience, attention, and perception. Phenomenological and embodied interaction scholarship (e.g., \cite{Dourish2001,Kirsh2013,Noe2004-NOEAIP,Varela2017EmbodiedMind}) emphasizes that interactive systems transform how people inhabit and interpret their bodies, not merely how they perform tasks. Somaesthetics and soma design further foreground cultivated bodily awareness, affect, and expression \cite{Hook2018}, offering approaches for designing systems that respect experiential qualities such as vulnerability, dignity, and comfort.

Body-centric computing and related perspectives \cite{benford-body-centric-dagstuhl-2018,mueller2018body,schraefel2018body} underscore how interactive systems become entangled with everyday bodily practices. Classic and contemporary work on wearability \cite{Gemperle_Kasabach_Stivoric_Bauer_Martin_1998,jabari-fashion} demonstrates that the material, physical, and social integration of body-centric computing directly shapes bodily experience. As emerging soft materials and computational garments show, wearability is not neutral---it determines comfort, compliance, social acceptability, and users’ sense of autonomy. However, this broader literature largely treats the body as a site of input or experience, rather than a target of actuation. This distinction is central to our paper: understanding bodily control requires building on these insights while expanding attention to how actuation may shape experience, agency, and perception beyond what interaction-oriented models capture.


\subsection{Technologies that Act on or Co-Act with the Body}

A wide range of technologies now intervene in or generate bodily movement. Electrical muscle stimulation (EMS) has demonstrated how computationally triggered contractions can guide, adjust, or override motion (e.g., \cite{Tamaki2011,kasahara_EMS_Speed_2021,knibbe_automaticEMS_2017,Knibbe2018}). Subsequent EMS work explored performance enhancement, haptic steering, safety mechanisms, and mixed-initiative movement guidance (e.g., \cite{Faltaous2020,Shahu2023EmsObstacleAvoidance,nith2024Splitbody}).

Beyond EMS, researchers have developed systems using pneumatic actuation (e.g., \cite{Saini_Pneuma_2024,Li:Playskin:2022, luo2022digital}), vibrotactile and haptic cues (e.g., \cite{kim2020defining,schneider2017haptic}), shape-changing interfaces and responsive materials (e.g., \cite{steed2021mechatronic, luo2022digital}), and galvanic vestibular stimulation (e.g., \cite{Maeda2005,Sra2019GVS,byrne2016BalanceNinja,byrne2020VertigoTochi}). \change{These technologies support a spectrum of control, ranging from subtle prompting that shapes, guides, or nudges bodily action while leaving execution and final decision-making with the user to strong physical interventions that generate bodily movement or override motor execution, making resistance difficult or impossible in the moment. Taken together, these prior works suggest that bodily control is not tied to a single technology but is emerging across diverse actuation strategies.}

Another stream of work investigates systems that co-move with users. Exoskeletons and prosthetic devices often combine sensing and actuation to support mobility, strength, or rehabilitation (e.g., \cite{herr2009exoskeletons,Sun2022,pons2010rehabilitation,Sun2022}). Furthermore, prior work demonstrates that computational systems can increasingly shape physical movement (e.g., \cite{Maeda2005,Sra2019GVS,byrne2016BalanceNinja,byrne2020VertigoTochi}). 
Related research on haptic guidance, rehabilitation robotics, and robotic gait support (e.g., \cite{koopman2014speed,Casadio2012}) shows how bodily coordination emerges from complex negotiation between human intentions and machine dynamics.

\change{
This view of bodily control systems as acting on or co-acting with the body resonates with perspectives from "Actor–Network Theory" by Latour \cite{Latour2005}, which conceptualize agency as distributed across networks of human and non-human actors, including bodily control systems. From these perspectives, such bodily control systems do not merely execute human intentions, but they can also shape, constrain, and redirect action in practice. This notion of systems "pushing back" highlights how bodily control systems should be seen as co-acting with users rather than functioning as neutral tools, providing a useful lens for understanding these emerging forms of bodily coordination.}

Taken together, we find that prior work on technologies that act on or co-act with the body is dispersed across fields, often focusing on technical or usability aspects in isolation. A unified conceptual understanding of bodily control – spanning experiential, technical, design, and ethical dimensions – remains largely underdeveloped. This motivates our effort to articulate grand challenges that can guide future research.

\subsection{Shared Control, Agency, and the Experience of Being Moved}

A third relevant thread concerns shared control between humans and computational systems. Foundational work in automation, mixed-initiative interaction, human-in-the-loop, and human-machine teaming (e.g., \cite{Abbink2018,Berberian:2019,Nunes:2018,chiou-trusting-automation,kerne-combinformation-mixed-initiative,FRAUNE2021102573}) 
shows that control is neither binary nor fixed but dynamically negotiated. Research on intelligibility and accountability (e.g., \cite{Bellotti01122001}) highlights the importance of making system logic perceptible when humans depend on algorithmic partners.

Empirical studies on sense of agency further demonstrate that the timing, predictability, and alignment of system actions critically shape whether people feel in control. This has been explored through input modalities (e.g., \cite{Coyle2012,bergstrom2018really,Bergstroem2022,wang2025handows,li2024investigating,li2025evaluting}), cognitive neuroscience (e.g., \cite{haggard2017sense,gallagher2000philosophical}), and computationally mediated bodily movement (e.g., \cite{padrao2016violating,Tajima2022}). These questions become especially salient when a system acts through the user's body. Even subtle system-driven motion can shift users’ perceived authorship, responsibility, or ownership of action \cite{Benford2021Contesting}. Research on embodied trajectories, looseness of control, and negotiated agency in bodily interaction, e.g., (\cite{Benford2021Contesting,MUELLER2021102643})
illustrates that control is experiential, interpretive, and relational, not purely technical.

Work on bodily diversity and critical disability studies (e.g., \cite{spiel-nothing-about-us-without-us,Spiel2021BodiesOfTEI,Britton2017,payne-ethically-engage-fat-people}) urges HCI to recognize that bodies are not uniform, and that many systems embed normative assumptions about strength, size, sensation, mobility, gender, and ability. Such insights are essential for bodily control systems, where calibration, force, comfort, and compliance are deeply entangled with bodily difference.

Across these areas, common themes emerge: maintaining agency, supporting transitions between system- and user-led action, avoiding overreliance, respecting bodily difference, and ensuring the legibility and safety of computational interventions. What remains missing is a consolidated articulation of the challenges shared across these themes. This paper brings these threads together to articulate grand challenges for designing computers’ control over our bodies.

\section{Methodology}
\label{sec:Method}

We conducted a workshop over 5~days with 24~international experts. 
This approach drew inspiration from previous HCI efforts that aimed to articulate grand challenges in other sub-fields of HCI~\cite{Elvitigala2024, Ens2021, Mueller2024WaterHCI}.

\begin{table*}[t]
\caption{Demographic information of workshop participants. Organizers marked with \dag. ``Exp.'' is experience.}
\label{tbl:Participants}
\centering
\begin{tabular}{lllrp{7.0cm}}
\toprule
\textbf{Role} & \makebox[6mm][r]{\textbf{Country}} & \textbf{Gender} & \textbf{Exp. (Years)} & \textbf{Research Background and Expertise} \\ 
\midrule
Prof \dag & UK & Woman & 6 & Affective  Computing; Assistive Technology  \\
Prof & DE & Woman & 3 & Wearable Computing; Multimedia \\
Industry & UK & Man & 4 & Interaction Design; Wearable Computing; Haptics \\
PhD & US & Man & 4 & Human-AI Interaction; Interaction Design; XR \\
Assistant Prof & AU &  Man & 10 & Interaction Design; Wearable Computing; Ethics \\
Industry \dag & US & Woman & 2 & Neuroscience; XR \\
Prof & JP & Man & 15 & Wearable Computing; XR \\
Associate Prof & AU & Man & 7 & Wearable Computing; XR; Interaction Design \\
Prof & UK & Man & 4 & Human-AI Interaction; XR \\
PhD & UK & Man & 2 & Interaction Design; XR; Embodied Interaction \\
Assistant Prof & CN & Woman & 6 & Wearable Computing; Haptics \\
Assistant Prof & UK & Man & 7 & Interaction Design; Wearable Computing \\
Industry & DK & Woman & 10 & Interaction Design; Play \\
Prof \dag & AU & Man & 10 & Interaction Design; Play; Wearable Computing\\
Post-Doc & UK & Woman & 7 & Embodied Cognition; Interaction Design \\
PhD & AU & Man & 8 & Interaction Design; Wearable Computing; Play \\
Assistant Prof & DK & Man & 3 & XR; Wearable Computing \\
Prof & UK & Woman & 9& Psychology; Embodied Interaction; Interaction Design \\
Prof & DE & Man & 4& Haptics; Interaction Design; XR \\
PhD & AU & Man & 7& Wearable Computing; Haptics \\
Associate Prof & CA & Man & 6& Interaction Design; Play; Haptics \\
PhD & AT & Woman & 5& Assistive Tech; Cognitive Science \\
Assistant Prof \dag & US & Woman & 8& Human-AI Interaction; XR; Haptics \\
Prof & DE & Man & 7& Wearable Computing; Haptics; Fabrication \\ 
\bottomrule
\end{tabular}
\end{table*}

\subsection{Participants}
The recruitment process aimed to gather a diverse group of experts by focusing on (a) international and institutional representation, (b) a range of expertise across research topics, and (c) varied career stages.
Our final 24~participants (Table~\ref{tbl:Participants}) reflects a diversity of perspectives, backgrounds, and research interests, spanning areas such as assistive technology, ethics, and haptic interfaces as well as varied epistemologies, from theory-driven to more hands-on design research.
The participant cohort was intentionally composed of experts spanning expertise across a range of application domains. 
This heterogeneity ensured that discussions drew from technical, experiential, and ethical expertise. 
As a result, the emerging challenges reflect both practical constraints (e.g., actuator feasibility, calibration difficulty) and experiential or societal concerns (e.g., agency, consent, bodily diversity). Participants’ disciplinary backgrounds directly shaped how topics were problematized. 
For example, assistive technology specialists foregrounded transitions from dependency to autonomy, while haptics researchers emphasized timing, latency, and intelligibility.

\begin{figure*}[t]
\centering
\includegraphics[width=\linewidth]{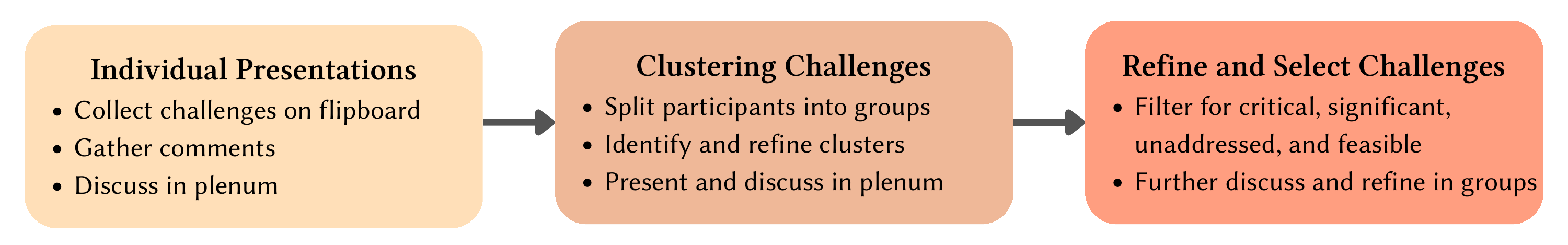}
\Description{The figure illustrates the three-stage process we used during the workshop to generate, cluster, and refine research challenges. First, participants individually presented their challenges, which were collected on flipboards and discussed in a plenary session. Next, participants were split into three groups to collaboratively identify and refine clusters of related challenges, which were then presented back to the plenary. Finally, participants collectively refined and selected challenges by filtering for those considered critical, significant, unaddressed, and feasible, followed by further in-depth group discussions.}
\caption{Workshop process for identifying and refining grand challenges}
\label{fig:Procedure}
\end{figure*}

\subsection{Procedure}

To arrive at the grand challenges, we used an incremental process where we first collected a large set of issues to discuss and then refined them in multiple stages (Figure~\ref{fig:Procedure}). 
The overall goal of articulating grand challenges and how previous workshops had approached this (see, e.g.,~\cite{Elvitigala2024, Ens2021, Mueller2024WaterHCI}) were introduced at the beginning of the workshop.
This informed the categories to explore, but also the processes to follow.

\subsubsection{Identification of Potential Challenges}
We began with each participant presenting their research, articulating the challenges they experienced over the years. 
We recorded these challenges on four A2 sheets during each presentation, clustering them initially under the categories technology, users, design, and ethics.
During this exercise, we also collected reasons why we would want to yield control over our bodies.
This in turn led to discussions about the different conceptual levels of the challenges faced in participants' investigations, calling for a consideration of the various theoretical perspectives through which an articulation of the grand challenges could be approached. 

\subsubsection{Collective Grouping and Discussion of Challenges}
The organizers encouraged participants to add comments to the sheets at any point during the sessions, noting challenges and opportunities that came to mind related to designing computers' control over our bodies. 
This process resulted in 118~challenges, which were recorded in a Google sheet and categorized under the initial categories, serving as a basis for further discussions.
Given space constraints, we provide the full set of raw challenge statements in the supplementary material. 
This offers transparency into how granular contributions progressed toward higher-level categories and allows readers to trace how early-stage insights informed the final grand challenges.

\subsubsection{Clustering Entries}
\begin{figure*}
\centering
\includegraphics[width=\linewidth]{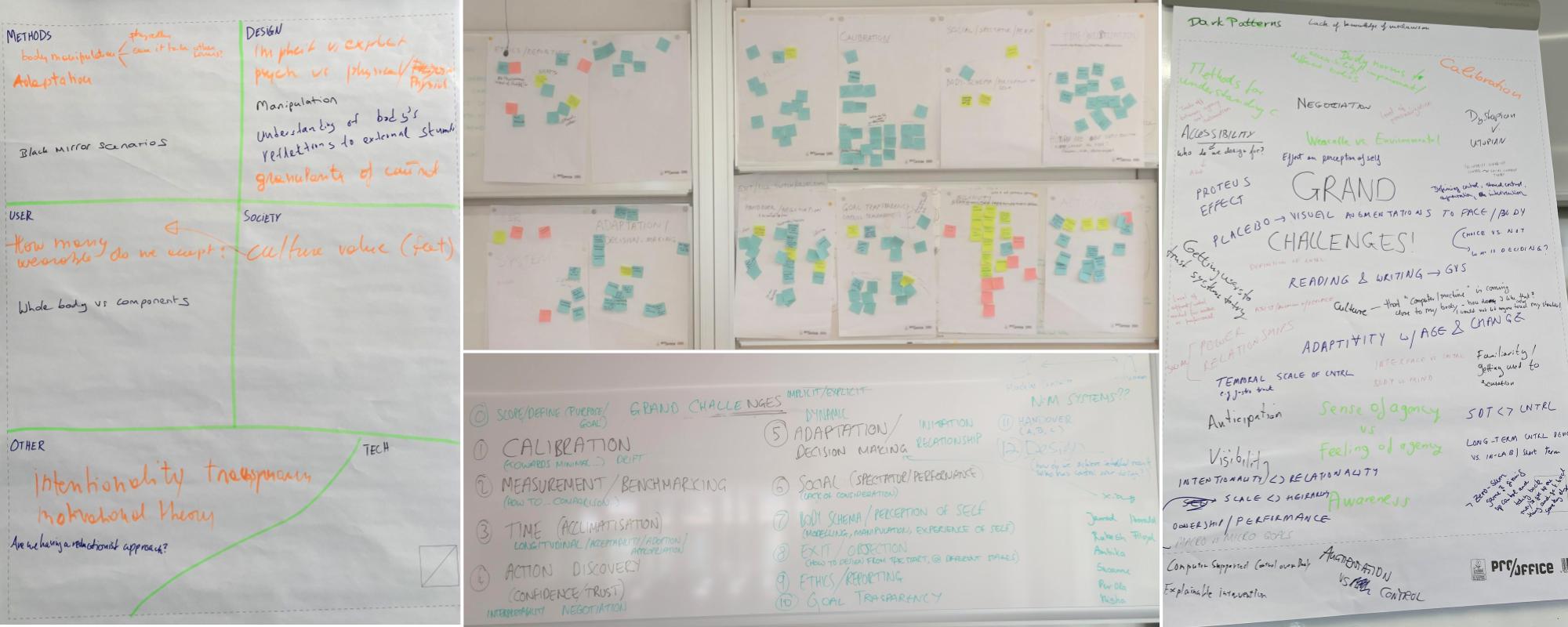}
\caption{Photographs of collaborative efforts, including whiteboards, A2 sheets, and post-its. The visuals capture thematic discussions on topics like adaptation, user perspectives, societal values, and technical challenges, organized into categories such as methods, design, user, technology and society.}
\Description{A collage of photographs showing whiteboards, A2 sheets, and post-its from a collaborative workshop. The visuals display grouped notes and outputs organized into categories like methods, design, user, society, and technology.}
\label{fig:Method}
\end{figure*}

In small groups, participants clustered the 118 entries into ten grand challenges (Figure~\ref{fig:Method}). 

Our approach followed previous HCI grand challenges efforts, as 3 participants had prior experience with these ~\cite{Mueller2024FoodGC,Mueller2020IntegrationNextSteps,Elvitigala2024}. This involved recording notes and photos from all whiteboards and post-it notes for each session and sharing them online for later reflection. We clustered the collected data in a collaborative and reflexive manner, where discussions evolved from practice and theory in an intertwined way, going back and forth between design examples and abstract knowledge. Our groupings started off from prior work that used technology, users, design, and ethics categories ~\cite{Mueller2024WaterHCI,Elvitigala2024,Ens2021}. As we wanted to let the participants’ experiences drive the emergence of individual challenges, we allowed the clustering of notes from the initial presentations to emerge organically. The organizers had experience with various qualitative research methods, so we acknowledge that their expertise might have tainted these efforts.

We opted not to use ``grounded theory'' because this requires researchers to be open-minded during data collection. 
The focus of the seminar was to gather expert insights, hence objectivity was traded for advanced knowledge. 
Furthermore, we were concerned that, given the diversity in our participant pool, that a formal approach like grounded theory would give stronger voice to those that were more experienced with this method than others. We wanted to be open in regard to what will come out of our open workshop process rather than guided by a specific research direction upfront, preventing the organizers (who prepared the workshop and hence started their preparations earlier) to have too much influence, for example, through a predetermined research direction arising from some of their grant projects at their respective home institutions. We hence focused on bottom-up sense-making while acknowledging everyone’ prior expertise. Since grand-challenge identification appears to benefit from participant domain knowledge ~\cite{Mueller2024FoodGC,Mueller2020IntegrationNextSteps,Elvitigala2024}, the structured but interpretive nature of our approach seemingly enabled to work productively with informed perspectives rather than treating them as bias to eliminate.
This approach to analyzing workshop content has previously been successful in producing grand challenge papers presented at major HCI publication venues~\cite{Mueller2020IntegrationNextSteps,Elvitigala2024}.
The resulting clusters were revisited and discussed the following day in a town hall setting to reach consensus and refine the details of each grand challenge.

\subsubsection{Group Discussion}
Through the clustering process, several important tensions emerged in relation to bodily control systems, including guidance vs autonomy, internal vs external control, dependency vs recovery, empowerment vs overreach, aesthetic vs social acceptance.  These discussions informed our efforts to consolidate the initial challenges before applying the following inclusion criteria, inspired by \citet{Elvitigala2024}: (1) Is the challenge uniquely significant or more pronounced in the context of designing computers' control over our body? (2) Is the challenge critical for the field to advance and not easily solved? (3) Has the challenge not been addressed yet in current work? (4) Is it feasible to solve within a decade? Using the post-it notes from the clustering, participants discussed which grand challenges met these criteria.

\subsubsection{Final Grouping and Selection of Grand Challenges}
Participants were divided into breakout groups to elaborate on proposed challenges. 
In a second round, groups were reshuffled to discuss different challenges. 
Following each round, the groups presented highlights in a plenary session: a brief description of each possibly revised challenge and its list of sub-challenges. 
We incrementally iterated these into our final grouping of grand challenges, aiming to reach consensus while capturing the major ideas that were discussed; this was done in concert with being aware (through reminders by the organizers) of the time constraints of the workshop format.

\subsubsection{Reflexivity and Negotiation of Perspectives}
To ensure that the resulting grand challenges reflected a range of positions rather than a consensus-by-smoothing, we attended to how differing views emerged and were handled throughout the workshop. During synthesis sessions, the organizers noted points of disagreement and encouraged participants to articulate competing interpretations. 
Rather than resolving divergences, we preserved them as analytic resources to understand tensions around bodily control systems. 
The final grand challenges therefore represent negotiated interpretations across diverse perspectives, rather than an averaged or unanimously agreed set of grand challenges.




\begin{figure}[t]
\centering
\includegraphics[width=\linewidth]{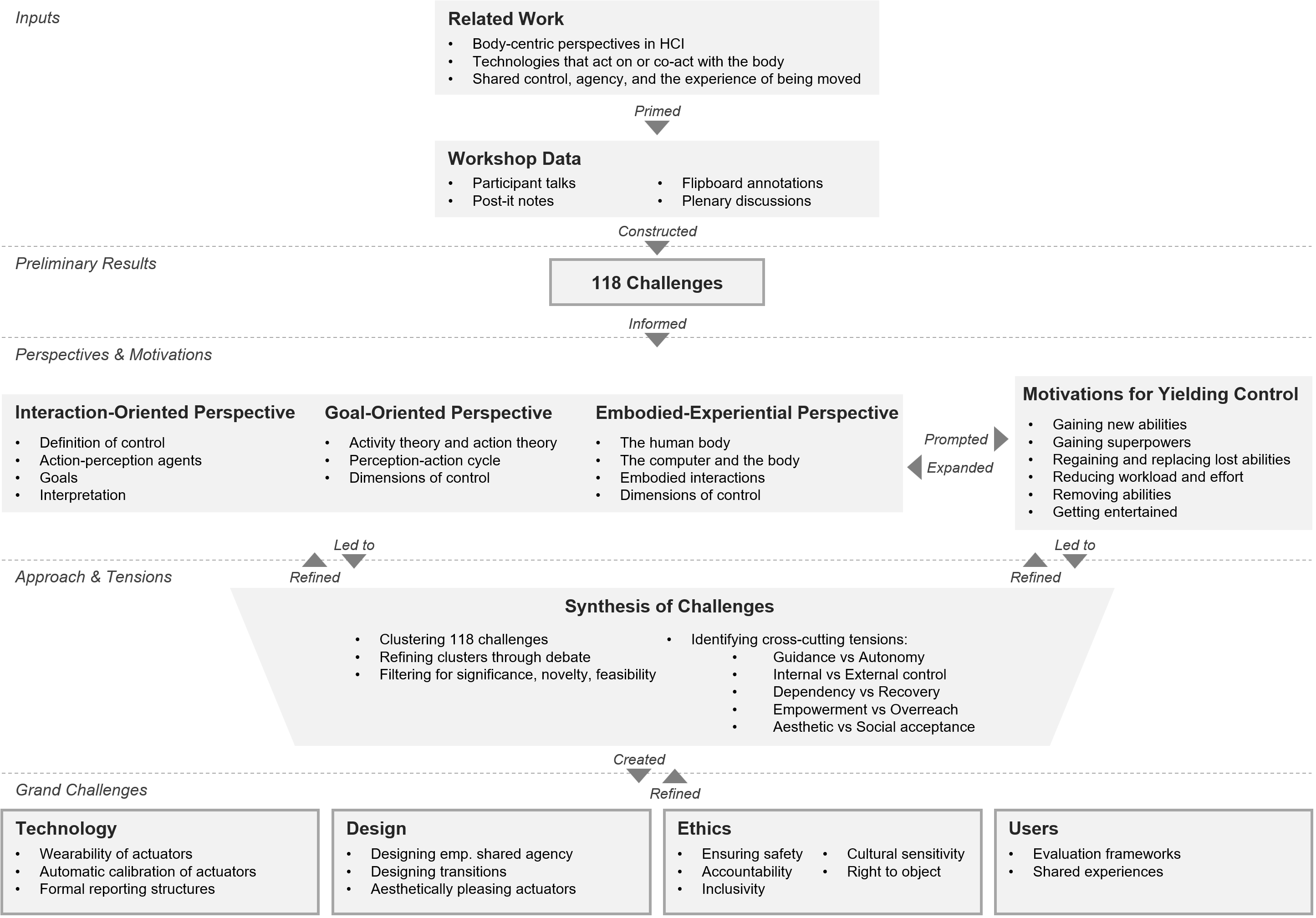}
\caption{From Inputs to Grand Challenges. }
\Description{Workflow divided into five regions stacked vertically.
First region, "Inputs," contains two rectangular blocks. The left block, "Related Work," lists: body-centric perspectives in HCI; technologies that act on or co-act with the body; shared control, agency, and the experience of being moved. The right block, "Workshop Data," lists: participant talks; post-it notes; flipboard annotations; plenary discussions. A downward arrow labeled "Primed" points to the next region. Second region, "Preliminary Results," shows a single centered box labeled "118 Challenges," with an arrow labeled "Constructed" above and "Informed" below. Third region, "Perspectives and Motivations," shows four horizontally arranged blocks. From left to right: Interaction-Oriented Perspective, listing: definition of control; action-perception agents; goals; interpretation. Goal-Oriented Perspective, listing: activity theory and action theory; perception-action cycle; dimensions of control. Embodied-Experiential Perspective, listing: the human body; the computer and the body; embodied interactions; dimensions of control. Motivations for Yielding Control, listing: gaining new abilities; gaining superpowers; regaining and replacing lost abilities; reducing workload and effort; removing abilities; getting entertained. Arrows on this row show "Prompted" pointing left toward the perspectives and "Expanded" pointing right toward motivations. Fourth region, "Approach and Tensions," shows a large trapezoidal block labeled "Synthesis of Challenges," containing: clustering 118 challenges; identifying cross-cutting tensions; refining clusters through debate; filtering for significance, novelty, and feasibility; and the following tensions: guidance versus autonomy; internal versus external control; dependency versus recovery; empowerment versus overreach; aesthetic versus social acceptance. Arrows labeled "Led to" appear above and below. Fifth region, "Grand Challenges," contains four rectangular blocks side by side. Technology lists: wearability of actuators; automatic calibration of actuators; formal reporting structures. Design lists: designing empirical shared agency; designing transitions; aesthetically pleasing actuators. Ethics lists: ensuring safety; accountability; inclusivity; cultural sensitivity; right to object. Users lists: evaluation frameworks; shared experiences. A downward arrow labeled "Refined" appears above this region, and an upward arrow labeled "Created" points back to the synthesis region.}
\label{fig:arrivingGCS}
\end{figure}

\subsection{Expansion}
\label{sec:MethodSummary}

Our process to determine the grand challenges around computers taking control of our bodies started from the participants' individual research perspectives.
Through subsequent activities, we refined, aggregated, and extended these challenges.
While the main outcomes are the grand challenges, the workshop also yielded other insights that underlie these challenges.
Specifically, we found that participant groups used different perspectives on control in their articulations of the challenges.
As these perspectives underlie the challenges, we describe these first in Section~\ref{sec:Perspectives}.
Furthermore, our discussions revealed that perspectives on the motivations for these control changes differed within the groups, too.
As these then also informed subsequent challenges, we describe them in Section~\ref{sec:Reasons}.
We then articulate the resulting grand challenges in Section~\ref{sec:GrandChallenges}.
Figure~\ref{fig:arrivingGCS} illustrates this process.

\section{Perspectives on Control over Bodies} 
\label{sec:Perspectives}

Our discussions revealed three complementary perspectives on computers controlling our bodies: \textit{interaction-oriented}, \textit{goal-oriented}, and  \textit{embodied-experiential}. 
Together, they form a spectrum of ways to help understand how computers can control our bodies. 
Each perspective emphasizes different dimensions of control, while collectively capturing the technical, experiential, and ethical complexities of bodily control. These perspectives apply across diverse actuation modalities, including electrical muscle stimulation, pneumatic systems, vibrotactile arrays, shape-changing materials, and robotic exoskeletons, and across varied user populations. Importantly, they highlight that control is dynamic, negotiated, and relational, rather than static.
This spectrum resonates with broader accounts of human--computer integration, such as Mueller et al.’s notion of \textit{intertwined} interactions \cite{mueller2023toward}, where bodies and computational processes mutually shape one another.
This spectrum has informed our investigation into the grand challenges (see Section~\ref{sec:GrandChallenges}), however, we believe that it might also have utility on their own.
For example, researchers focused on low-level interaction patterns (interaction-oriented) might find that expanding their view to task-oriented action loops (goal-oriented) enriches their understanding of how interactions manifest in real-world applications. Furthermore, we note that our perspectives complement each other, rather than compete with one another.


\subsection{Interaction-Oriented Perspective}

This perspective focuses on low-level perception-action loops using formal engineering terminology and connects to representations of system implementations. 
It provides vocabulary and formal reasoning for understanding control, including how intentions, commands, and outcomes align. This perspective takes inspiration from control theory (e.g., motor control in humans or robots) as well as cognitive science (e.g., Norman’s activity loop \cite{norman2013design}). 
Interaction-oriented reasoning is particularly useful for analyzing precision, timing, and reliability of bodily control systems. 
It also surfaces technical tensions such as guidance versus autonomy and internal versus externalized control. 
However, it does not fully capture users’ lived experience or social context, and assumes normative body characteristics, highlighting the need for complementary perspectives.

\begin{figure}
    \centering
    \includegraphics[width=\linewidth]{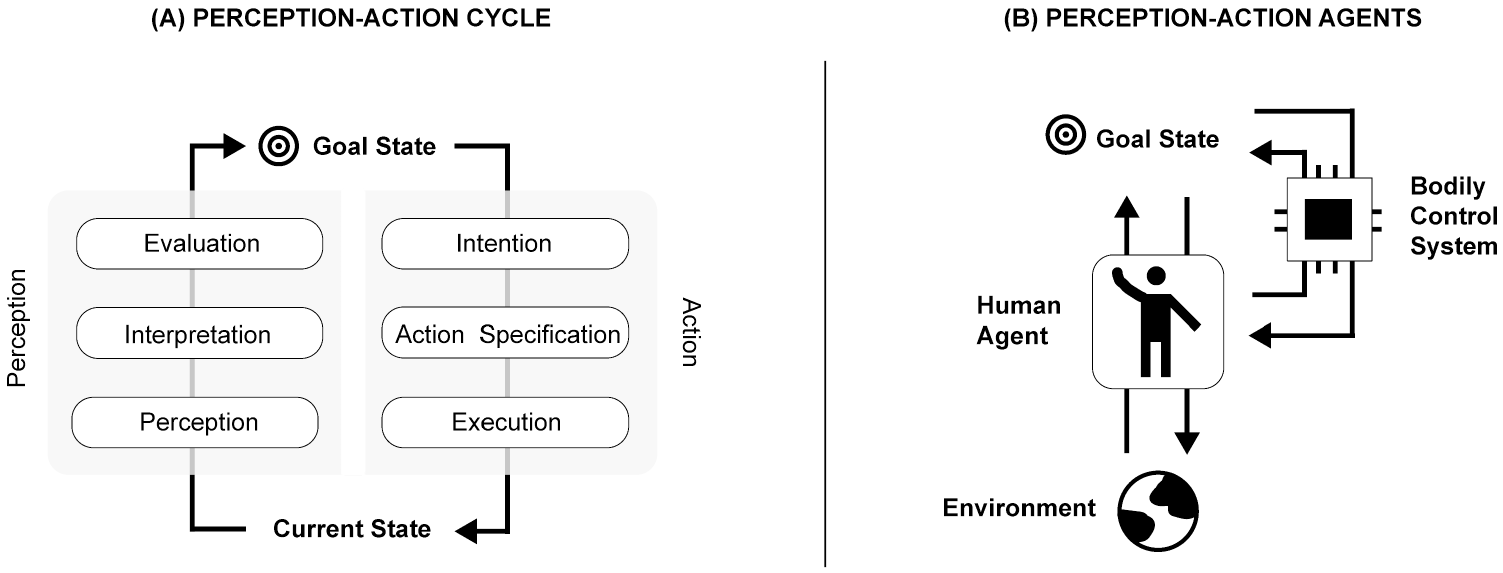}
    \caption{Interaction-Oriented Perspective: (A) Goals lead to intentions, action specifications, and motor execution. The resulting changes in the environment generate sensory feedback that is perceived, interpreted, and evaluated against the goal, completing the perception–action cycle. (B) Humans, computational systems, and body-control agents each run their own perception–action loops. These loops interact through shared states and exchanged signals in a dynamic multi-agent system.}
    \Description{Panel A, titled "Perception–Action Cycle," presents a closed loop between a Goal State at the top and a Current State at the bottom. Two vertically arranged shaded areas appear side by side. The right area is labeled "Action" and contains three stacked blocks in the following order from top to bottom: Intention, Action Specification, and Execution. The left area is labeled "Perception" and contains three stacked blocks in the following order from bottom to top: Perception, Interpretation, and Evaluation. The layout depicts information flowing from the Goal State through the Action blocks to the Current State, then from the Current State through the Perception blocks back to the Goal State, forming a recurring cycle. Panel B, titled "Perception–Action Agents," shows the same Goal State icon at the top along with two continuous feedback loops. One loop connects the Goal State and the Environment, represented by a globe, and passes through a human figure labeled Human Agent. The other loop connects the Goal State and the Human Agent through a box marked Bodily Control Systems, represented by a chip icon.}
    \label{fig:InteractionOriented}
\end{figure}

\subsubsection{Control as the ability to change state}

Control is the ability to change state to achieve a goal. To reach a goal, we need several bodily processes to align (see Gallagher~\cite{gallagher2000philosophical}). 
First, we have the intention of reaching a goal. 
Intentions provide the motivation and plan for acting, while command is the order to execute an intention. 
For an action to align with a goal, the outcome must be interpreted both at the perceptual and higher cognitive levels as successful. 
Successful execution of the described procedure to reach a desired goal will change one state to another and enable control.

\subsubsection{Action-Perception Agents}
Control is a system-level property. 
A system is made up of actors, from one to many. Every actor can constitute an action-perception loop in itself. As such, a human and a computer can constitute individual actors. 
An action-perception loop is defined as the combination of a set of possible actions that can be performed on the world, together with senses that can perceive the world. 
This way, our control system is part of defining our reality \cite{gonzalez2017model}.
For a user, actions are functions of biomechanics, for example, motor actions for moving limbs or breath control for producing speech. 
Digital system actions may include functions of displays or speakers, and now also actuators that control the human body. 
In both cases, systems are formed by interconnected modules that can fail with undesirable results or succeed and reach desired goals. 
An outcome fails if the actions do not align with the perceptions. 
That happens when actions are based on insufficient or misinterpreted sensory information, or if there is no information to construct an intention (planning) and command to act \cite{gallagher2000philosophical}. Figure~\ref{fig:InteractionOriented} shows how bodily control systems interact with humans in a combined action-perception cycle with the environment.

\subsubsection{Goals}
A goal is an actor's desired state. Goals exist at multiple levels of granularity, from micro goals (i.e., make the hand grasp) to macro goals (make the user fit).
They can be more traditional task-focused goals (e.g., ``do-goals'' \cite{hassenzahl2010needs}) or more human-focused motives or needs (e.g., ``be-goals'', like belonging or competence \cite{hassenzahl2010needs}). 
Micro goals are mostly more accessible to capture and measure (e.g., "Did the EMS-controlled hand grasp the object?"). 
Macro goals are made up of micro-goals, but these may not necessarily be easy to decompose. 


Interestingly, with the advancements of AI, we see that our traditional command-control paradigms are seemingly shifting more into goal-oriented computing. Most AI systems are currently designed as ``prompt and repeat'': if a prompt does not result in the desired outcome, the user has to ask for the task to be performed again in a different way (re-prompt)~\cite{zamfirescu2023johnny}. This might require intensive prompting and re-prompting, which might not be feasible in control scenarios like driving.

\subsubsection{Interpretation}
A system must interpret the data it receives to determine a given action. 
This involves multiple layers of interpretation. 
At the lowest control-state level, the system is pre-interpretive and sub-perceptual (e.g., controlling a grasp). 
At the second level, individual actor interpretations arise (e.g., ``is the grasp strong enough?'').
At the third level, a social reflection across all actors in a system arises (e.g., ``are we all holding together?''). 
A further layer of meta-reflection also exists, including societal reflections, spanning time, ethics, etc. (e.g., ``is this grasping ethical?'').
Prior work highlighted that humans are judged by their intentions while computers are judged by the outcomes \cite{hidalgo2021humans}. 
This means that if we want a computer to retain the intentions of the human, the user will need to participate throughout the whole action.

\subsection{Goal-Oriented Perspective} 

The \textit{goal-oriented perspective} focuses on how users, systems, and environments interact to accomplish specific tasks or objectives. 
This perspective considers shared and adaptive control, illustrating moments where system autonomy may conflict with user intention, and highlights the dynamic negotiation of control over time. 
It also situates bodily control within broader activity contexts, social norms, and environmental constraints.
The relationship between the user, the system, and the world is key to achieving the goal.

For instance, a user who is rehabilitating from a limb injury might want a system to temporarily take control over that limb whilst they recover. Another example could be a user who wants to learn a new skill and allows a system to take control over their body to help teach them this new skill. Alternatively, users may require longer-term interventions, where systems are required to empower or extend a user’s existing abilities, such as in the case of restoring some level of mobility to a mobility-impaired individual.

\subsubsection{Activity Theory and Action Theory}

During the workshop, activity theory was often mentioned as a way to approach the goal-oriented perspective. Activity Theory \cite{engestrom1999perspectives} frames user actions within a broader context of purposeful activities, emphasizing how computers as tools mediate between users and their objectives. Activity Theory's hierarchical structure of activities, actions, and operations offers a framework to analyze how users employ systems to achieve specific goals. By highlighting the dynamic nature of tool-mediated activities, Activity Theory can help account for how system-controlled interactions evolve, adapting to changing goals and contexts.

Through the lens of Action Theory \cite{engestrom1999perspectives}, the goal-oriented perspective can illuminate insights for designing systems where computers take control over a user's body, such as exoskeletons in physical rehabilitation. By framing these technologies as mediating tools within goal-oriented activities, designers can better understand how to integrate them into a user's broader objectives and contexts. This perspective encourages consideration of how control is dynamically shared between the user and system, adapting to changing goals and physical capabilities throughout the rehabilitation process. This perspective also asks designers to consider the social and environmental factors that might influence the acceptance and effective use of such systems, potentially leading to more holistic and user-centered designs that align closely with therapeutic goals.

\subsubsection{Perception-Action Cycle}

The goal-oriented perspective emphasizes the intricate interplay between the user, the system, and the environment, recognizing these elements as forming a dynamic triangle of interaction~\cite{Jagacinski2018Control}. This dynamic triangle is shaped by continuous feedback loops of perception and action. The system must discern not just how to help the user, but when and to what extent, allowing for user-initiated requests for assistance even when the system might deem them unnecessary. This raises interesting control questions. For example, what if a user requests walking assistance from a system that controls the legs, but the system believes the user should walk independently due to being overweight?

The goal-oriented perspective considers varied temporal aspects, from short, repetitive tasks to long-term, persistent use scenarios, such as injury recovery. The perception-action cycle serves as the backbone of this interaction, enabling adaptive responses based on real-time analysis. This cooperative approach fosters user agency that can build trust (important as users need to trust systems that take control as otherwise their muscles might tighten up, hindering computer-control) as the system continuously adapts to support users in achieving their goals within the given environmental context.

\subsubsection{Dimensions of Control}
The dimensions of control that emerge from the goal-oriented perspective reflect a complex and dynamic relationship between the user and the system. Control is not a fixed attribute, but a fluid entity, continuously negotiated based on task progress, user confidence, and environmental factors. 
While goal-oriented systems might be complex to design when it comes to bodily control, one important caveat of goal-oriented systems is that they will perform on well-defined goals, but might struggle to generalize or be multipurpose.

\begin{figure*}[t]
\centering
\includegraphics[width=\linewidth]{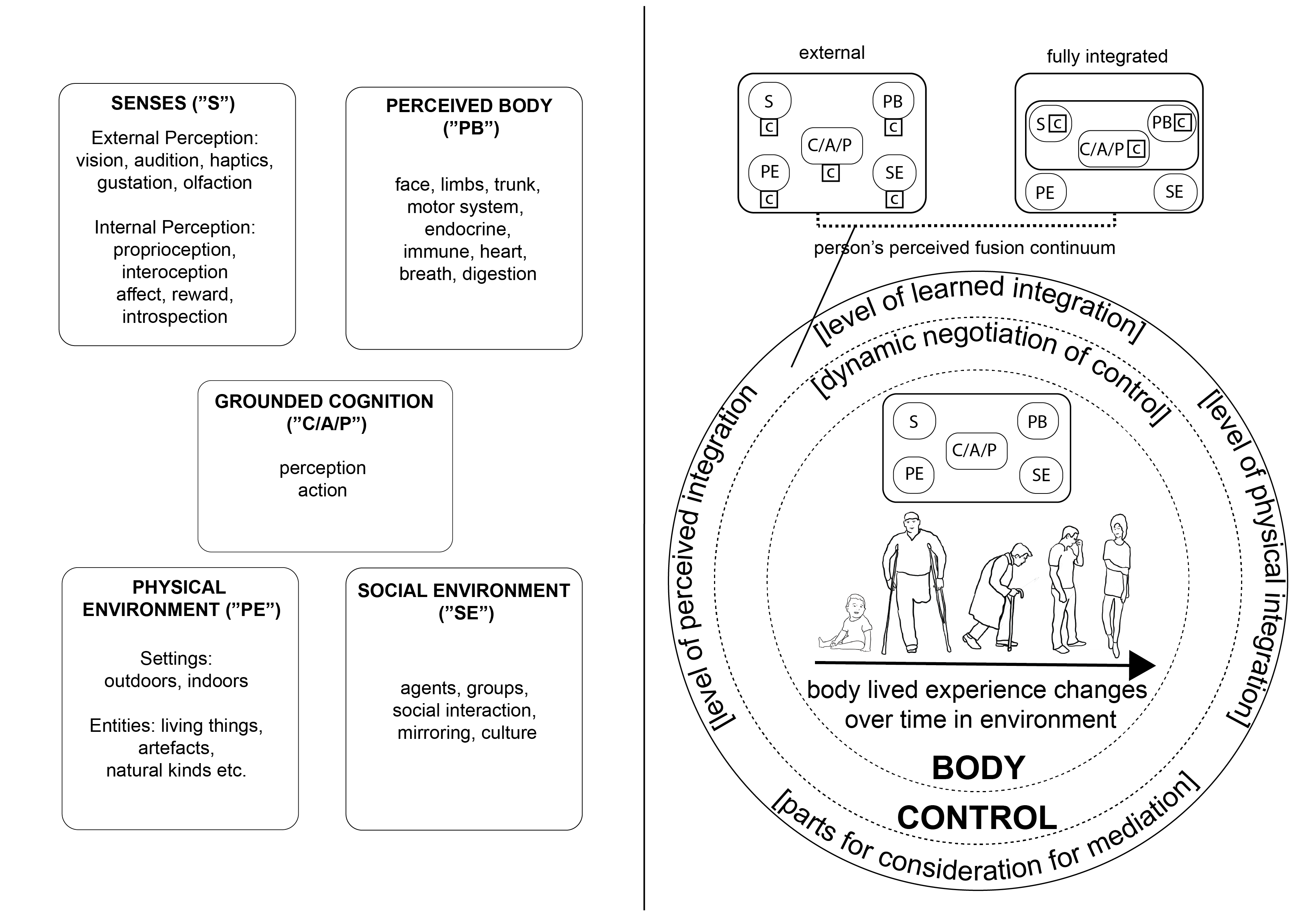}
\caption{Domains of cognition from a grounded perspective foregrounding cognition as perception-action coupling. Adapted from~\cite[p. 2]{Barsalou2020}. Left: Cognition emerges on a continuum of perception-action couplings (C/A/P) in mechanisms of senses (S), perceived body (PB), physical environment (PE) and social environment (SE). Right: the diagram illustrates the body-computer relationship mediated by dimensions of control leading to a dynamic negotiation around control. The C in the small square indicates the position of the computer with respect to the human from being external to fully integrated.}
\Description{Diagram showing domains of cognition from a grounded perspective, emphasizing perception-action couplings. The left section categorizes cognition into five domains: 'Senses (S)' including external and internal perception; 'Perceived Body (PB)' referring to bodily systems like motor and digestive; 'Grounded Cognition (C/A/P)' focusing on perception and action; 'Physical Environment (PE)' describing settings and entities; and 'Social Environment (SE)' covering agents, interactions, and culture. The right section illustrates the evolving relationship between the body and computer, transitioning from external to fully integrated. It highlights levels of physical and perceived integration, learned integration, and dynamic control negotiation over time, influenced by environmental interactions. Symbols and diagrams represent fusion and control dynamics across life stages.}
\label{fig:ControlGraph}
\end{figure*}

\subsection{Embodied-experiential perspective}

The embodied-experiential perspective focuses on how bodies dynamically negotiate control amid their various relationships, with more emphasis on social, embodied, and societal considerations~\cite{10.1145/3364998}. 
It centers the lived experience of the human body, foregrounding perception-action coupling, bodily diversity, and social context. 
It emphasizes the subjective perception of agency, trust, and consent, highlighting that control is fluid and relational. 
By incorporating considerations of diverse body types, abilities, and cultural norms, this perspective challenges able-bodied-centric assumptions and guides inclusive design.
The body and computer engage in a dynamic relational interaction. 
Rather than on implementation, this perspective focuses primarily on experience. 
It draws on a grounded and enactive cognition framing of human experience (e.g., \cite{Barsalou2020, Varela2017EmbodiedMind, Noe2004-NOEAIP}) where experience unfolds on a perception-action continuum.
Here, bodily and environmental processes are understood to exist as an ``ongoing coupling'' (e.g., \cite[p. 31]{articleDiPaolo}) foregrounding relationality. 
For example, Li and Kristensson present an on-body curved-selection technique \cite{li2025bend} that uses the forearm as a proprioceptive anchor to structure and disambiguate interaction within dense virtual environments.
This perspective develops three aspects: the human body, the computer, and dimensions of control (Figure~\ref{fig:ControlGraph}).

\subsubsection{The human body} 

The body --- and the embodied nature of experience --- is central to a person's experience of control. 
First, we draw on notions of embodied cognition (e.g.,~\cite{Barsalou2020}) that foreground human experience arising through sensorimotor engagement with control that takes place in distinct, yet interrelated domains: sensory, body, physical environment and social environment. 
Second, we foreground the diversity of bodies to encompass development, disability, age, transition, etc., to illustrate how bodies can differ (e.g., \cite{bourdieu1990logic,haimson-trans-tech,toups-dugas-pole-2024}), and that this difference re-frames their relationship with the notion of control and computers. 
Finally. the notion of lived experience describes how bodies change, develop, and evolve in the moment of interaction and beyond (e.g., \cite{Laz2003,haimson-trans-tech,cardenas-dragon,toups-dugas-pole-2024}).

It is important to note that considering bodily diversity in HCI still has a long way to go, which will ultimately also aid computers in controlling human bodies. For example, in the workshop, we discussed how sensor-equipped earbuds do not work in larger ears, heart rate monitors struggle with darker skin tones, step counters fail to reflect wheelchair users’ exercise, most smartwatches do not work for people who have arms with physical differences, and many wearables’ form factors do not suit the changing bodies of gender-transitioning individuals. 

\change{
Other examples are EMS systems that produce uneven or excessive contractions across bodies with different muscle mass, fat distribution, or injury histories, complicating calibration and safe force delivery. Similarly, exoskeletons often assume normative gait patterns, limb symmetry, or strength recovery trajectories, which can result in uncomfortable, ineffective, or even harmful assistance for users with atypical movement patterns or asymmetrical bodies.}

Furthermore, we also acknowledge that our own work often involves testing prototypes only with users having normative bodies. By inviting people with bodies outside these norms, such as people with disabilities, injuries (permanent and temporary), growing bodies (children/teenagers), aging bodies (seniors), and other non-normative bodies (including people from the LGBTQIA+ community), we could become better designers. We still have more work to do in strengthening our knowledge about how assumptions about bodily norms influence our design; this enhanced knowledge will only benefit our understanding of how to design computers that take control of our bodies.

\subsubsection{The computer and the body} 

From an embodiment perspective, it is important to consider the following: (1) The form: The shape and the location of any attachment to the human body have implications for how the human body is, acts, and becomes changed or extended, and acts on itself and the environment. (2) Sensors: What the computer can sense from the environment and the human. (3) Actuators: \textit{How} the computer implements control over the human body, and the granularity of control that it exerts. (4) Human-machine interface: how the level of integration of the computer within the human body is implemented as this will affect how it is processed and integrated within the embodied experience.

\subsubsection{Embodied interactions}

Embodied interactions~\cite{Dourish2001} describe interactions that take place within the world, where the world plays a part in the interaction. 
For computers that control the body, the world becomes important not only to the user but also to the system \cite{Bellotti01122001}.
For example, a computer can (e.g., algorithms, AI), through the use of appropriate sensors, continually monitor the world around the user and adapt accordingly. 
This means that, if controlling a person's walking gait, the computer can speed up, slow down, or send a command to change direction if it senses an obstacle in the path. 
Furthermore, the computer could react to input from the user (perhaps they want to override the suggestion, or if the user's heart rate has dangerously risen, the system needs to slow the user down). 
Another example is a user who wishes to improve their cardiovascular health by connecting a heart rate monitor to a treadmill that can automatically change speed to keep the user's heart beating at the required beats per minute. 
The goal is to improve cardiovascular health: the system monitors the heart rate thanks to the heart rate sensor, and reacts to cause the user to run faster or slower by speeding up or slowing down the treadmill motor. This completes the action-perception loop. 

\subsubsection{Dimensions of control} 

From an embodiment perspective, foregrounding perception and action in the context of this paper (e.g., \cite{Barsalou2020, Varela2017EmbodiedMind, Noe2004-NOEAIP}), control could then be conceptualized as being dynamically negotiated between the body and the computer as part of an ongoing dialogic process where both entities shape control during the interaction. A dynamic negotiation process between the two entities may expose various dimensions of control emerging from the situational aspects, the designed function and present functionalities, or user(s) of the experience. By considering the negotiation as an ongoing process during interaction, these dimensions may adopt "fluctuating" qualities of control during the interaction experience. By this we mean that control may undergo changes in relation to perception and action, rather than remain static. 
Control between computer and human is therefore a dynamic negotiation, where control lies and is modulated as a process within the ongoing coupling between human (here: body) and the environment (here: computer). 
The perception of control between the computer and the body may not always align. 
The computer may explicitly control a body, yet the user may not be aware of this control, or the user may be aware of the control, but is resistant to this. 
The fluidity of control between the two can be captured by 4 dimensions of control: (1) Agency: the degree to which the user has control or perceives they are in control (of the body). (2) The directness of control: whether control is implicit or explicit.
(3) Initiation of control: who initiates a change in control.
(4) Transparency of control: whether there is visibility or invisibility of control. 

Together, these perspectives provide a holistic understanding: interaction-oriented reasoning informs technical feasibility; goal-oriented reasoning informs task-level design and negotiation of control; embodied-experiential reasoning informs user experience, ethics, and social acceptability. Recognizing the limitations and assumptions of each perspective is crucial. While each provides a valuable lens, none fully captures all aspects of bodily control in isolation. Integrating the three helps to identify challenges across technology, design, users, and ethics, and provides guidance for developing systems that are technically robust, socially acceptable, and ethically responsible.

\section{Why Yield Control over our Bodies?}
\label{sec:Reasons}


In addition to the perspectives discussed in the previous section, our analysis surfaced several recurring motivations for why people might yield control of their bodies to computers. These motivations originated from different combinations of technical necessity, task-oriented goals, and embodied experience. They also illuminate how the tensions outlined earlier – such as guidance versus autonomy, internal versus externalized control, and support versus dependency – play out in practice. Here, we present these motivations not as universal categories, but as lenses for understanding why bodily control may be desired, accepted, or resisted across different bodies, contexts, and technologies.

\subsection{Gaining New Abilities}

Designing computers to control our bodies opens up the possibility of gaining new abilities, both physical and cognitive. For example, EMS can enable users to draw slip streamlines around a car model without needing to understand the underlying physics of airflow \cite{Lopes2016MusclePlotter}. In this case, the computer takes control of the user's hand, allowing them to perform a task they would otherwise be unable to do. However, it is important to distinguish between merely performing a skill and actually acquiring it. While such systems can enhance task performance --- improving completion time and reducing workload --- they may not necessarily result in long-term learning or skill retention \cite{Shahu2023Skillab}.

Looking to the future, computers may also enhance mental processes through brain stimulation techniques, such as transcranial alternating current stimulation (tACS). Research is only just beginning to reveal how such stimulation could be used to control cognitive functions \cite{Hermann:tacs:2018, Paulus:2016}. This suggests a future where computers not only guide our physical actions but also directly control our mental processes.

\subsection{Gaining Superpowers}

Enabling computers to control our bodies also offers the potential to enhance human abilities to ``superhuman'' levels \cite{superhumansports_2019}, akin to comic book superheroes \cite{Liu:Superpower:2024}. Metaphors of superpowers have long inspired various HCI fields, such as information visualization \cite{willett2021perception}, and can similarly motivate the design of body-controlling systems. For example, researchers have used EMS to trigger muscle movements, allowing for reaction times faster than humanly possible \cite{kasahara_EMS_Speed_2021}. This approach enables tasks such as pressing a camera shutter button on time for high-speed photography, which would be challenging without the system \cite{lopes:preemptive:2019}.

\subsection{Regaining and Replacing Lost Abilities}

Enabling computers to control our bodies can offer the potential to restore abilities lost due to disease, injury, disability, or ageing. 
Physical therapy with the help of robotics is a prominent example, such as an active upper-limb exoskeleton that helps with physical training \cite{Cheng:2022}. 
On-body systems can be used for children with upper extremity movement impairments in their activities of daily living \cite{Li:Playskin:2022}. 
In these examples, an active interface directly on one's body controls movements to enable the user to eventually regain control. 
If the system had pre-injury data, such as motion capture from walking, an advanced system could help the user not just regain mobility but walk in their unique style (similar to a walking `fingerprint' used for identification). 
The goal of such systems is often to create a training effect, allowing users to recover and eventually no longer need assistance. 
Thus, the system acts as a temporary aid during recovery until the user regains full ability.

\subsection{Reducing Workload and Effort}

Bodily control systems can also help offload some of the workload associated with heavy or complex physical activities \cite{exo_rehab_sixian_2021}. For example, early navigation systems in cars aimed to minimize the time needed to shift between watching the road and checking the navigation display by visually guiding the driver's eyes back to the relevant point on the map \cite{gazemarks:2010:schmidt}. Extending this concept further, computers can control or modulate bodily functions and actions to assist users in focusing on essential tasks and information \cite{Basil1994, Longo2022, nith2024Splitbody}. For example, robotic exoskeletons have been developed to aid in physical rehabilitation by providing external control over limb movements, thereby reducing the effort required by the user and optimizing performance \cite{herr2009exoskeletons, pons2010rehabilitation}. In instances where multitasking, cognitive, or motor activities incur higher cognitive load and degrade performance, letting computers take control of some of the tasks or body parts can help maintain focus and performance on other tasks \cite{koopman2014speed, liu2022real}. For example, users can benefit from delegating the control of one of their hands to a system, reducing the sensorimotor load in exchange for some of their autonomy and freeing sensorimotor resources for other needs \cite{nith2024Splitbody}.

\subsection{Removing Abilities}

Computers that control our bodies can also restrict or remove our ability to move, with several potential use cases.

First, they can prevent unsafe actions. For example, in industrial settings, untrained or inattentive workers may face hazards like harmful materials or dangerous machinery. Body control systems can prevent such exposure, enhancing safety \cite{butler2017exoskeleton}. This approach is particularly useful in collaborative robotics, allowing full use of robotic capabilities without risk of injury \cite{weckenborg2022harmonizing}. In complex environments, systems that actively take control, such as automated lane-keeping systems that turn the steering wheel, can prevent accidents, supplementing traditional alarms \cite{Stratmann:ShipBridges:2019}. Second, bodily control can support behavior change, potentially aiding in treating addictions or compulsive behaviors \cite{Park2015Dynamic}. For users motivated to change but struggling to do so, a computer might help by physically guiding or preventing specific actions. Third, removing abilities can be useful in simulations and entertainment, such as simulating impaired mobility to build empathy. For example, the TeslaSuit (EMS suit) can temporarily disrupt lower limb function, simulating leg injuries \cite{fitzgibbon2010shared}. Finally, we must acknowledge that bodily control systems could be misused to control others. For example, electronic tags already monitor offenders, and future technologies may be used for more nuanced control, not only within legal and penal systems.

\subsection{Getting Entertained}

Another reason for yielding control to computers is for entertainment. 
Early game research emphasized the importance of player control (e.g., the GameFlow model \cite{10.1145/1077246.1077253}). 
However, recent work has revealed the potential of manipulating control as a design resource in bodily games (e.g., see \cite{floyd2021limited}). 
This can lead to novel and intriguing bodily experiences. 
A classic example is a roller coaster: people relinquish control, yielding to a pre-determined playful thrill experience ~\cite{Marshall2011}. Recent HCI research has explored this design space. For example, physical games that impose constraints can challenge players to move within limitations \cite{10.1145/3411764.3445622, 10.1145/3334480.3381652}. 
Similarly, Galvanic Vestibular Stimulation (GVS) technology can be used to induce vertigo that is perceived positively in games \cite{byrne2016BalanceNinja, byrne2020VertigoTochi}. 
These examples suggest that yielding bodily control can be a powerful tool for creating unique entertainment experiences.




\section{Grand Challenges}
\label{sec:GrandChallenges}

Our grand challenges are presented around four themes: technology, design, ethics, and users, oriented on prior grand challenge work in other HCI sub-fields~\cite{Mueller2024WaterHCI,Elvitigala2024,Mueller2024FoodGC, Mueller2020IntegrationNextSteps}.

\subsection{Technology}

Technological grand challenges concern practicalities of such systems, but also their configuration and specification.

\subsubsection{Wearability of actuators}
Bodily control systems face significant challenges in ergonomic wearability. 
Actuators’ \textit{weight} often correlates with power; high-force actuators are typically heavy, leading to user fatigue. 
Another crucial factor is \textit{rigidity}. It is well-known that technologies that are too rigid in their material reduce wearability~\cite{Mueller2020IntegrationNextSteps, Gemperle_Kasabach_Stivoric_Bauer_Martin_1998}. 
Yet, we note some correlation between an actuator's strength and rigidity: not only are most actuators rigid (see mechanical exoskeletons), but the more forceful they are, the more rigid they need to be. 
Rigid and oftentimes rather bulky actuators stand in contrast to the ``softness'' of the human body, leading to reduced \textit{body conformity} and possibly even interfering with the user's motor activities.
Additionally, powerful actuators generate more heat, compromising \textit{thermal comfort} (or even \textit{safety}). 
This underscores the need for advances in softer, lightweight actuating technologies that still deliver sufficient force for body control.  
Advances around integrating electronics into textiles~\cite{Yu2024IrOnTex} and orthoses~\cite{Wang2023ThermoFit} point to potential ways to bring about such improved actuation.
Developers of actuators designed to control our bodies should also consider how users can mount and dismount them on the body.
For example, if a user is impaired or if the actuator is located in a hard-to-reach body location, this may introduce dependence on others to mount or dismount the equipment ~\cite{knibbe_skillsleeves}. 
This might mean that, from an interaction-oriented perspective, the system is performing correctly, however, the user is unable to achieve their goal, as from an embodied-experiential perspective, we see that the user requires help from another person, which works against the notion of independence that the system promises.
In this line, the person is using the system in a social context, for which we need to anticipate potential negative feedback from peers or bystanders of user’s wearing and using the technology in their day to day activities \cite{Koelle:unacceptable:2019} and how this affects social acceptability \cite{Koelle:SocialAcceptability:2020} of the system. 
We note that the boundary between tactile feedback and bodily movement stimulation is increasingly blurring thanks to technical advances, affecting how we see the wearability of actuators. For example, work on creating spatialized vibrotactile feedback systems (e.g.,~\cite{huang2025vibraforge}) suggests that previously considered “tactile” techniques can increasingly approach forms of guided movement. We believe that the progressive miniaturization and increased spatial resolution of tactile systems could be used to address the “stimulus/control” boundary and complement current efforts, such as those using EMS, blurring distinctions between notifications, guidance, nudging, and bodily control. Furthermore, technical advances may blur the boundaries between existing approaches to bodily control. For example, if exoskeletons become small enough to be worn discreetly under clothing, they could rival EMS in terms of concealability. Likewise, as vibrotactile arrays become more widely available, they may offer alternatives that compete with the accessibility of EMS devices. Furthermore, while wearability is still a challenge, many research labs have made significant progress on materials, form factors, and donning/doffing workflows~\cite{Yu2024IrOnTex, Wang2023ThermoFit}.

\subsubsection{Automatic calibration of actuators}
Bodily control systems are still very cumbersome to use as they require calibration to work for each user's body in the best possible way, and automatic calibration systems for actuators are still in their infancy. 
Calibration for personal factors like body weight, height, and skin thickness are often necessary. 
This is complex, as it may involve sensing specific physiological signals or targeting specific muscles, which are not easily accessible, as the goal-oriented perspective highlights~\cite{knibbe_automaticEMS_2017, Gange_Knibbe_2021}. 
Accurate ongoing calibration is essential but time-consuming, impacting user experience. Recalibration may be required during use due to body state or environmental changes.
Identifying the optimal time as to when and how the system needs to be recalibrated is also essential. 
Designing for minimal calibration is a highly important and challenging task~\cite{Gange_Knibbe_2021}. 
Promising directions include: leveraging closed-box optimization techniques~\cite{Gange_Knibbe_2021}; building on user's informed decisions to ease semi-automatic interactive calibration~\cite{Pohl2018EMSCalibration}; utilizing anatomical models to predict muscle activations~\cite{Baier2024Simulation}; and using closed-loop feedback from the same modality (e.g., EMS stimulation-EMG measurement~\cite{knibbe_automaticEMS_2017}).

\subsubsection{Formal reporting structures}
The interaction-oriented perspective highlights that understanding a system’s technical capacity for controlling the human body can be crucial. Formal reporting structures need to be developed to outline these capacities accurately. For example, if a system is used to assist in uphill movement, its performance (and limitations) must be known to ensure optimal use. This is especially important for actuators that initiate or take over motor functions. The lack of formal standards and structures for such technical reporting presents a significant challenge. Better formal reporting structures can support the development of prototyping tools and, in turn, substantially aid the design of technical solutions.

\subsection{Design}

Grand challenges around the design of control shifts concern shared agency, transitions between human and computer control, and aesthetics.

\subsubsection{Designing Empowering Shared Agency}

Sharing agency with a computer can be empowering; for instance, when an exoskeleton enables a user to lift an object they could not manage alone. Yet designing such shared agency remains difficult, and we currently lack actionable design knowledge for creating empowering systems. Advancing this area will require collaboration among user experience researchers, technologists, and designers. Prior work on human-in-the-loop  \cite{Nunes:2018,Berberian:2019} and mixed-initiative systems ~\cite{kerne-combinformation-mixed-initiative} offers useful starting points, but these approaches have largely focused on traditional interfaces such as mouse and keyboard input (with exceptions \cite{FRAUNE2021102573}). It remains unclear whether these interaction-oriented perspectives fully translate to contexts where computers act directly on the user’s body.

%

An analogy of car driving might be helpful here: yielding control over our body can be similar to the shift from manually steering a car to autonomous vehicles.
Between full and no control, the distinction between user-driven and system-driven actions is blurred. 
A computer braking faster than a human can react may feel empowering in an emergency. Yet this sense of empowerment can quickly break down. Such systems often assume the user would have made the same choice – an assumption that holds when a pedestrian steps into the road, but not necessarily when a small animal appears. Some drivers would brake; others would avoid braking to prevent being rear-ended. If the system brakes when the user would not, the mismatch can diminish perceived agency and erode trust (in addition to introducing additional ethical tensions). A similar tension can arise when an exoskeleton unexpectedly adjusts movement – for example, applying force when a person intends to gently lift a child because it interprets the action as part of a workout – can also undermine agency. These examples illustrate how misaligned guidance and autonomy can compromise shared agency, and why designers must carefully anticipate the limits of empowerment in such systems.

Designers need to balance control between the user and the system while ensuring that users feel empowered (embodied-experiential perspective) even when the system takes over.
The challenge is maintaining a sense of shared agency, ensuring system actions align with user intentions without feeling disempowering.
How to design for equality, interpretability, and predictability are important considerations.
Users should feel like collaborative partners and as in charge of what is happening~\cite{Coyle2012}.
Concepts such as ``intentional binding'' might be helpful for understanding such experiences~\cite{bergstrom2018really}, but designing for shared agency remains a significant challenge.

There are several ways designers could begin addressing this challenge, whether the goal is to distribute control between user and system more evenly or to allow one to take the lead when appropriate. A key next step would be developing instruments that measure a user’s perceived level of control in real time. Such metrics could enable systems to adjust their behavior more precisely and responsively than is currently possible.

\subsubsection{Designing Transitions between Human and System Control}

With transitions being an important aspect of interactions with computers~\cite{benford2009interaction, Benford2021Contesting}, the transitions between human and computer control also require attention.
Ceding control to the computer goes counter to user's baseline experience of mostly being in control of their bodies.
The result can be jarring and thus a smooth physical and psychological transition is important. 
This starts with the on-boarding phase, where users might need to cede control for the very first time, and ends with the off-boarding phase, where users have to return to no longer sharing any control.
Hence, seamless (see, e.g., \cite{chalmers2005gaming}) transitions are generally desirable and relevant across short- and long-term use. 

An example might be a scenario where a user wears an exoskeleton to assist with walking after an injury. 
Initially, the system might take full control to help the user regain mobility. As the user recovers, they will gradually take back control, but this process should be managed delicately to avoid abrupt changes that could cause discomfort or even injury. 
The system must anticipate the user's intentions and adjust its level of control accordingly, all while taking the user’s gradual recovery and hence improving abilities into account. 

This rehabilitation example illustrates that transitions between human and system control are not merely technical – they involve experiential and ethical tensions. Supportive control can gradually become dependency. How might a system detect such dependency and taper assistance to avoid overreliance? And if the system decides to reduce support, even in the user’s best interest, it still overrides the user, potentially reducing their sense of agency.

Designers therefore need to consider how users can feel they are regaining agency as recovery progresses. One question is whether involving a third party – such as a physiotherapist – might support this process. If so, how should systems manage shifting distributions of agency among the user, the system, and the therapist, especially given that users and systems interact moment-to-moment while therapists intervene only intermittently?
Or should the system reduce its level of control and adopt a physiotherapist-like role, providing a reassuring sense of presence in case something goes wrong, helping users overcome emotional barriers to agency, and gradually fading that presence as confidence grows? Alternatively, should it decrease its control while still signaling that support is available, but in a way that becomes increasingly subtle and less perceptible over time?

Future systems could benefit from advanced sensing and decision-making capabilities that could be driven by artificial intelligence that was trained with data from other people and the user's own past data to provide a personalized transition experience. 
However, such data is still scarce, and if it exists, knowledge on how to utilize it to facilitate such seamless transitions is limited.

User trust depends on perceiving transitions as supportive, with confidence in both the system’s assistance and their ability to regain control. This should allow gradual adjustment but also immediate full control in emergencies (e.g., via dedicated controls~\cite{Patibanda2022Fused}). 
Trust-building is probably essential for the adoption of such systems, yet can be difficult to achieve. 
For example, sudden or unexpected shifts in control can lead to a breakdown of trust, causing users to abandon a system. 
Yet, in safety-relevant environments, for example, there may only be very little time to inform the user of an emergency. 
Research on trust in automation \cite{chiou-trusting-automation,momen-trust-tesla} will need to account for systems that address control of human bodies. 

Software that gradually adjusts the level of control (e.g., by monitoring the user’s physical and cognitive state) could support the seamlessness of transitions.
However, how to design such closed-loop (see, e.g., \cite{Semertzidis:Dozer:2023,Kosmyna:AttentiveUI:2019}) transitions is relatively underexplored. 
Additionally, the challenge is to ensure that the user remains aware of the system's actions and intentions during these transitions to prevent disorientation or an unintended loss of agency. 
Prior work on awareness of EMS signals~\cite{patibanda2023auto} could inform such developments. 
Furthermore, it is important that users can easily discover~\cite{norman2010natural} that they have the power to (re-)gain control. 
However, more research from such an embodied-experiential perspective is needed to understand how to facilitate awareness and discoverability at the right level and at the right time. 

A useful next step would be to conduct a bodily control-centric review across the many disciplines that study shared control – such as engineering, computer science, and the social sciences – to clarify the field’s current state and trajectory. Such a synthesis could help identify gaps and prevent duplication of effort.

\subsubsection{Aesthetically pleasing actuators across state changes}

Another grand challenge from the embodied-experiential perspective is that we do not yet know how to design the actuators aesthetically. 
Aesthetics for any wearable system is an important factor influencing its acceptability ~\cite{weigel2015iskin,knibbe_skillsleeves,jabari-fashion}. 
Here, we extend this by pointing out that actuators worn on the body face an additional challenge: they do not only need to look aesthetically pleasing from a visual perspective \cite{jabari-fashion}, they need to look (and feel) pleasing across all their state changes (e.g., fully inflated and deflated pneumatic systems~\cite{Saini_Pneuma_2024}). 
Additionally, preferences for aesthetics evolve with fashion trends and individual tastes, impacting user acceptance.
We note that aesthetic appeal and social acceptance do not always align. A fashionable, high-end device may be visually appealing in some communities yet socially awkward in others, for example, see the history of Google Glass. Furthermore, a bodily control system designed to resemble an everyday object – such as a wristband – may be less aesthetically appealing due to its materials or construction, but its familiar form and location can make it far more socially acceptable and easier to integrate into daily life. A possible next step would be to organize a workshop with fashion experts, where they advise on designing aesthetically pleasing systems while technology experts account for the constraints of bodily control technologies.

\change{Future work would benefit from more systematic design exploration and evaluation. For example, comparative studies could examine how different distributions of control affect perceived agency, trust, and comfort across tasks and time. Longitudinal studies could investigate how users’ experiences of shared agency evolve as systems adapt, fade, or withdraw control. Developing and validating instruments to measure perceived agency and trust in bodily control systems would further enable designers to iteratively refine such systems.}


\subsection{Ethics}
The embodied-experiential perspective highlights that, with an increasing number of control systems, there is a need to address significant ethical challenges. Researchers, designers, industry practitioners, policymakers, etc. will need to work together to ensure that the development of these technologies is guided by ethical considerations. In general, it is necessary to include many groups to achieve successful inclusive designs \cite{spiel-nothing-about-us-without-us,spiel-criptopias} and be guided by principles of consent \cite{strengers-consent}.

\subsubsection{Ensuring safety}
Using computers that take control over the human body can introduce critical safety issues that must be addressed to ensure user protection and minimize negative impacts. 
Especially the interaction-oriented perspective highlights that we have only a limited understanding of how to design these systems safely. 
On the one hand, a computer taking control can increase safety, such as when directing a user away from danger and/or towards help in emergency situations~\cite{Faltaous2020, lopes222neckactuation}. 
On the other hand, the system's direct interaction with human physiology can lead to unforeseen risks, including unintended reflexive actions that could be dangerous. 
Some technologies, such as EMS, are known to potentially cause muscle strain and even burns, requiring robust safety features to prevent overuse~\cite{patibanda2023auto}.
Clear guidance and training for users are also essential for safe and effective operation \cite{Kono_Takahashi_Nakamura_Miyaki_Rekimoto_2018,Patibanda2022Fused}.

Technologies like exoskeletons should include fail-safes to prevent mechanical failures that could result in accidents. 
Ongoing user education can help mitigate risks. Concepts like passivity from engineering control, used in teleoperation and robotics, may help keep systems stable or allow human control during safety-critical moments.

Deploying these systems can impact the environment, including people and animals, which must remain safe and uncontaminated. Such systems should not compromise environmental integrity or create hazards. Users may also attempt to use these systems in unsuitable environments, like the shower, which could create electrical risks. Prior work has noted the importance of considering non-standard environments in HCI~\cite{Mueller2024WaterHCI}. Comprehensive safety protocols and rigorous testing are needed before user exposure to ensure safety. Ensuring the safety of these systems is key to their broader acceptance and long-term viability.

Furthermore, negative health effects can arise from over-reliance on a control system. For example, when using an exoskeleton to lift heavy items, users may gradually increase the weight they attempt to carry. If their perception of weight does not adjust accordingly, they might misjudge the load, risking dropped items and potential injury. Users could also experience cognitive overload if the system requires them to perform too many tasks simultaneously~\cite{nith2024Splitbody}. 
Moreover, systems that alter natural movement patterns may exacerbate existing health issues or introduce new ones. Psychological dependence on these systems is another concern, as users may experience withdrawal symptoms if access is lost. The long-term health impacts are not yet fully understood, highlighting the need for further research. We recommend implementing safeguards to ensure that the benefits of these systems do not compromise user well-being.

\subsubsection{Accountability}
The embodied-experiential perspective highlights that, as computers take control over the human body, accountability becomes an important ethical challenge. 
When a system exhibits autonomy, the line of responsibility can blur when failures occur \cite{killer-robot}. 
For instance, if an EMS system malfunctions and causes a user's arm to move and injure someone, who is responsible? Is it the user, the EMS hardware engineer, the software developer, the user experience designer, or the manufacturer? 
This issue becomes more complex as systems become interconnected and dependent on external systems (such as data processing centers in the cloud) and involve multiple stakeholders (developers, manufacturers, third-party services providers, etc.).
Users may grow increasingly reliant on these systems, but who is responsible for maintenance, updates, and safety might be unclear. 
Additionally, significant risks arise when a company discontinues support (e.g.,~\cite{strickland-bionic-eye}). 
This raises ethical concerns about abandonment and the need for regulatory oversight to ensure continued maintenance and accountability. One way to investigate accountability is through design fiction~\cite{speculative2013dunne}, using imagined systems to explore how accountability can break down. Such speculative scenarios can help designers identify vulnerabilities and develop more accountable control systems, similar to how research on dark patterns~\cite{greenberg2014dark, Liu:Superpower:2024} reveals harmful design strategies to inform better practices.

\subsubsection{Inclusivity}

If computers take control over our body, concerns arise due to the variability in the effectiveness and accessibility of the underlying technology, which depends on the user’s bodily characteristics. 
For instance, people vary in how they perceive physical stimuli \cite{li2021individual}, which might influence how they respond to actuators mounted on the body. 
Furthermore, these mounts are usually not very adaptable to different body shapes and sizes outside a normative range, which exclude many body types~\cite{Spiel2021BodiesOfTEI}. 
Age, gender, and other physical differences could also influence how these systems actuate individuals, which could lead to systems inadvertently favoring certain bodies over others.
Additionally, the controlling software might have assumptions about cultural, gender, or physical norms that may inadvertently exclude or disadvantage certain groups \cite{payne-ethically-engage-fat-people,keyes-misgendering}. 
Hence, we highlight the need for inclusive design practices that account for the wide variability in human bodies, ensuring that these systems do not exacerbate existing disparities but rather are accessible and effective for all users regardless of their bodily differences. One next step towards addressing this challenge could be to write a manifesto that requires signing researchers to commit to having study participants with non-normative bodies in their participant pool. However, such efforts might need to be accompanied by specific expertise and training requirements, depending on the individual cases.

\subsubsection{Cultural sensitivity}

Computers taking control presents a grand challenge in addressing cultural factors, as these systems may interact with deeply ingrained cultural practices and beliefs about who controls one's body. 
Therefore, the design and deployment of such systems must consider the diverse cultural contexts in which they will be used.
In particular, this needs to recognize that practices such as those related to body autonomy, privacy, modesty, and personal boundaries vary widely across different cultures, highlighted by the embodied-experiential perspective. 
For example, systems that require the removal of clothing to achieve direct skin contact~\cite{Patibanda2022Fused,patibanda2023auto,SBF_Patibanda_2024} may conflict with cultural norms regarding modesty or privacy, potentially alienating or excluding certain user groups. 
Moreover, the predominance of Western cultural perspectives in interaction design~\cite{cannanure2021decolonizing,kumar2018hci} could lead to the creation of systems that fail to accommodate the needs and values of non-Western users whose acceptance of a computer taking control of their body might be significantly different. 
A next step to address this challenge could be to conduct user studies across labs in order to gain insights into cultural differences when participants yield control to computers.

\subsubsection{Right to object}

One key consideration from the goal-oriented perspective is the possibility of objection and rejection; user consent is essential \cite{strengers-consent}. 
How can we design systems that do not force individuals to allow computers to control their bodies? 
This is especially important for vulnerable individuals, who may not want to be ``fixed'' or have their behavior normalized by external control. 
As such, the concept of normality and the body must be re-examined in the context of systems like exoskeletons~\cite{butnaru:2024}.  

A relatively easy first step might be to implement a kill-switch for every control system designers create in order to give users the opportunity to opt-out being controlled by the computer (see, e.g.,~\cite{wang2025lucientry}).

The question of who holds the power to decide whether we allow such control is crucial and social or economic pressures may push individuals into using these systems, for instance, to work longer hours or improve productivity.  
This includes the need for a deeper understanding of how consent and systemic pressures shape interactions with bodily control systems. For example, prior work has described systems that move a user’s hands away from the keyboard to encourage regular typing breaks~\cite{Saini_Pneuma_2024}. While well-intentioned, such a system could be misused: an employer might override break intervals by citing looming deadlines, or workers might feel pressured to disable the system entirely, undermining its purpose. Similar dynamics arise in healthcare. A teenager may rely on a gait-correcting exoskeleton to support long-term posture, yet social stigma at school could discourage its use. In both cases, external pressures can conflict with personal intentions, highlighting the need for designs that account for the social environments in which bodily control technologies operate.
We can also envision bodily control systems designed to prevent violent actions, yet such systems could be hacked or misused by governments or institutions for oppression. How can we ensure that users retain the right and ability to object?

Another concern is the potential for deskilling, when certain tasks, actions, or movements are offloaded to the computer, alleviating physical strain but potentially hindering our ability to learn new skills and diminishing our existing abilities.

\change{Future studies could examine how users negotiate consent, objection, and trust in real-world contexts where social or institutional pressures are present, such as workplaces or healthcare settings. Speculative and participatory methods, such as design fiction or scenario-based studies, could further help surface ethical breakdowns before deployment. Long-term field studies may be particularly valuable for understanding how ethical concerns such as dependency, coercion, and accountability emerge over time.}

\subsection{Users}

Questions around the evaluation of systems that control users' bodies as well as multi-user scenarios constitute this set of grand challenges.

\subsubsection{Evaluation frameworks}

One challenge is the lack of an understanding of what makes a successful control system, speaking to the embodied-experiential perspective. 
The task is, therefore, to develop evaluation frameworks for assessing the user experience. 
This challenge is particularly pressing given the complexity of interactions between technology and human physiology.
For example, the diversity and complexity of actuation technology means evaluation is a major challenge \cite{schneider2017haptic}, with standardized, specific evaluation tools being critically needed \cite{kim2020defining}. 
The core questions (based on prior work around the need for evaluation frameworks in other HCI areas~\cite{Ens2021}) revolve around understanding why, when, and how to evaluate these experiences effectively.

First, the question of \textit{why} we need to evaluate such experiences is crucial. Is the goal to demonstrate that these systems can enhance human capabilities, such as improving mobility or rehabilitation? Or is it to ensure that the integration of these systems with the human body does not lead to negative outcomes, such as discomfort, dependency, or loss of autonomy? These varied purposes highlight the need for flexible and multi-faceted evaluation frameworks that can adapt to the specific goals of different technologies and their applications. We believe that understanding the underlying purpose of evaluation is essential for framing relevant and meaningful studies.

The \textit{when} of evaluation is equally challenging. 
Real-time evaluation offers the potential for capturing immediate, visceral reactions to the computer taking control, which can be useful for understanding how these systems interact with the human body and are perceived in dynamic environments. 
However, conducting evaluations during active use can be invasive or impractical, such as when jogging with an EMS ~\cite{hassan2017footstriker}. 
For instance, asking a user to provide feedback while they are navigating a complex terrain might disrupt their concentration and compromise safety. 
On the other hand, post-use evaluations may miss transient physiological or emotional states that occur while being controlled.
This also relies on the user’s memory, which can be imperfect, particularly in recalling specific physical sensations or emotional reactions~\cite{overdevest2023towards}.

Understanding \textit{who} the users are and \textit{where} they use these technologies is another important challenge. 
The diversity of potential users, from patients in rehabilitation to athletes seeking performance enhancement, asks for tailored evaluation approaches that consider each group's specific needs, abilities, goals, and contexts. 
For instance, elderly users might require different evaluation metrics compared to younger, more physically capable users. 
Additionally, the context in which these technologies are used – whether in clinical settings, homes, or outdoor environments – can impact the evaluation process. 
Clinical settings offer controlled conditions, but may not capture real-world challenges, while evaluations in naturalistic environments can reveal everyday use issues but introduce confounding variables and measurement issues.

Another crucial aspect is understanding \textit{how} to evaluate these experiences, particularly in the context of long-term use. 
While short-term studies are valuable for identifying immediate effects and usability issues, they generally fall short when it comes to providing insights into any long-term implications, such as whether the user attributes the actuator's power as their own, saying \textit{``I did that''} instead of \textit{``The computer did this for me''} \cite{Coyle2012}. 
Therefore, we believe that evaluating these systems over extended periods is required to ensure that they do not lead to unintended negative outcomes. However, defining what constitutes ``long-term'' is itself a challenge. 

There are also practical challenges related to \textit{what} specific features or outcomes should be evaluated. 
Given the intimate connection that can exist between the system and the human body, traditional evaluation methods might not be sufficient. 
Researchers need to develop new tools and methodologies that can capture both quantitative and qualitative data on user experiences, physiological responses, and psychological effects. 
For instance, the evaluation of exoskeletons that take control might benefit from sensors to monitor muscle activity, joint stress, and user fatigue in real-time, as well as surveys and interviews to assess a sense of agency~\cite{Coyle2012}, feeling of agency~\cite{Bergstroem2022}, and awareness of control~\cite{Wegner1999}.
Similarly, evaluating EMS systems might involve tracking muscle recovery rates, pain levels, and overall physical performance alongside user-reported outcomes such as motivation, well-being, and acceptance \cite{Faltaous2020,Faltaous2024emsacceptance, Shahu2022EMSAccept}.

\subsubsection{Shared experiences}
Engaging with computers that take control is often a deeply personal experience, yet it can also be a social one, as the embodied-experiential perspective reminds us. For example, prior work reported that participants, unprompted by the study researchers, used a control system around EMS with other people to create novel social experiences: they connected the sensor to one person and the electrode to another, resulting in shared laughter~\cite{patibanda2023auto}.
This suggests that there is potential for bodily control systems to become social. However, designing systems where control over the user’s body is not only shared with the computer but also with other people remains largely underexplored. This gap in design knowledge includes both implementation details and experiential aspects. For example, little is known about how to support users in social activities where control is dynamically shared between the computer and other users, which could result in vastly different experiences.

The appeal of social interaction might push users to attempt actions beyond their system’s safe limits, similar to how social dynamics can lead exertion game players to engage in activities that they might refrain from when alone~\cite{mueller2011designing,mueller2014movement}. 
Prior HCI research has explored the use of technology to balance exertion activities where participants have different physical abilities ~\cite{altimira2017enhancing,altimira2016digitally}. 
When applied to control systems, this concept becomes more complex, as misalignment or miscommunication could lead to disengagement, system abandonment, or even injury.

\section{Limitations and Future Work}
We acknowledge our work’s limitations and use them to articulate potential avenues for future work. Many of these limitations are shared with other studies aiming to outline grand challenges in other specific HCI subfields~\cite{Mueller2024WaterHCI,Elvitigala2024,Mueller2024FoodGC,Mueller2020IntegrationNextSteps,Ens2021}.

Articulating grand challenges through workshops is common~\cite{Mueller2024FoodGC,Elvitigala2024}, but there is no definitive validation proving its effectiveness. This includes considerations such as workshop length. While some previous workshops, like ours, lasted five days ~\cite{Mueller2024FoodGC,Mueller2020IntegrationNextSteps,Elvitigala2024} others were as short as one day ~\cite{Mueller2024WaterHCI}. Extending the workshop to multiple days may limit participant diversity due to personal or institutional constraints. However, the longer duration allowed for more in-depth discussions compared to shorter formats. Our workshop required travel and accommodation on-site fostering intense collaboration. Our participants reported strong rapport, which we believe enhanced the quality of the discussions and, consequently, the article. However, we acknowledge that the requirement for travel and accommodation costs limited the diversity of our participant pool. 
Because the workshop venue required an invited-participant format, organizers had to assemble a list of potential attendees. This enabled long-term planning and helped ensure the workshop’s success, but it also introduced bias: participation was limited to individuals the organizers knew or could identify online, which may have influenced the resulting grand challenges. Even so, we hope this initial set of challenges offers a valuable starting point for others to refine and expand.

Furthermore, we acknowledge that our workshop could have considered more diverse participants (incl. people with a less "positive mindset", although we note that each of our breakout groups had discussions around ethics, dystopian futures and dark patterns). Prior work in HCI has already pointed to the benefits of considering undesirable futures (looking beyond positive effects)~\cite{greenberg2014dark}. Hence, we suggest that future work investigates how to incorporate such efforts. We also note that our approach of handling different views via giving everyone write access to a shared online document could have led to a consensus approach in which divergent views could have become diluted through repeated deleting, editing and rewriting, resulting in attenuated statements that might lose its rhetoric strength in shaping the future of the field. Future workshops could also include experts from physiology, physiotherapy, rehabilitation, sports science, ethics, policy making, arts, etc. (although we note that our workshop participants had many years of experience working within such application areas in close interaction with both experts and users, such as patients and athletes). Although we followed group sizes used in similar prior work, further validation could determine the optimal size ~\cite{Mueller2024FoodGC,Mueller2020IntegrationNextSteps,Elvitigala2024,Ens2021}.  Prior papers on other bodily HCI aspects have used the same recruitment focus and group size, suggesting that it might also offer value here ~\cite{Mueller2024FoodGC,Mueller2020IntegrationNextSteps,Elvitigala2024}.

Future work could also investigate methods for validating the effectiveness of grand challenge articulations. For instance, research teams working on bodily control technologies could be divided, with some informed about the grand challenges and others not, to assess if prior knowledge of these challenges is beneficial. We also point out that we see our work not as an end but a starting point to be discussed, refuted and refined in the future.
\section{Conclusion}
To move the field forward as a whole, ultimately aiming to help produce a more humane technology future, we have described a set of grand challenges that researchers face when aiming to design computers that take control over our bodies. These challenges are inherently interwoven, both conceptually and technically, necessitating concerted, collaborative endeavors for resolution. 
By confronting the challenges expounded in this article, the collective capacity of HCI will be propelled toward a more comprehensive realization of this sub-field's potential. This paper serves as a commitment to that endeavor.

\begin{acks}
The authors thank Schloss Dagstuhl for their support.
Joe Marshall was supported by the \grantsponsor{UKRI1}{UKRI Trustworthy Autonomous Systems Hub}{https://gow.epsrc.ukri.org/NGBOViewGrant.aspx?GrantRef=EP/V00784X/1} under grant  \grantnum{UKRI1}{EP/V00784X/1}. 
Minna Nygren and Nadia Berthouze were supported by the \grantsponsor{UKRI2}{UKRI From Sensing to Collaboration programme grant}{https://gtr.ukri.org/projects?ref=EP\%2FV000748\%2F1&pn=0&fetchSize=25&selectedSortableField=parentPublicationTitle&selectedSortOrder=ASC} under grant \grantnum{UKRI2}{EP/V000748/1}. 
Xiang Li was supported by the China Scholarship Council (CSC) International Cambridge Scholarship (No. 202208320092). Per Ola Kristensson was supported by the EPSRC (grant EP/W02456X/1). Florian `Floyd' Mueller thanks the Australian Research Council, especially DP190102068, DP200102612 and LP210200656.
\end{acks}

\bibliographystyle{ACM-Reference-Format}
\bibliography{Bibliography}

\end{document}